\documentclass[a4paper,11pt]{article}
\usepackage[margin=1in]{geometry}
\usepackage[usenames,dvipsnames,table]{xcolor}
\usepackage{multirow}
\usepackage{makecell}
\usepackage{pdflscape}
\usepackage{graphicx}
\usepackage{cite}
\usepackage{amssymb,amsmath,mathrsfs}
\usepackage{subfig}
\usepackage{authblk}
\usepackage[colorlinks=true,urlcolor=blue
,anchorcolor=blue,citecolor=blue,filecolor=blue,linkcolor=red,menucolor=blue
]{hyperref}
\usepackage{verbatim}
\usepackage{authblk}
\usepackage{xspace}

\usepackage{relsize}
\def\babar{\mbox{\slshape B\kern-0.1em{\smaller A}\kern-0.1em
    B\kern-0.1em{\smaller A\kern-0.2em R}}}

\newcommand{\GeV}{\mathrm{GeV}}
\newcommand{\MeV}{\mathrm{MeV}}

\newcommand{\beq}{\begin{equation}}
\newcommand{\eeq}{\end{equation}}

\newcommand{\OSM}{\mathcal{O}_{\mathrm{SM}}}

\DeclareMathOperator\tr{tr}

\newcommand{\madgraph}{\texttt{MadGraph}\xspace}
\newcommand{\pythia}{\texttt{Pythia}\xspace}
\newcommand{\fb}{\mathrm{fb}}
\newcommand{\ab}{\mathrm{ab}}
\newcommand{\cm}{\mathrm{cm}}
\newcommand{\Ndv}{N_\mathrm{dv}}
\newcommand{\Ntr}{N_\mathrm{tr}}

\newcommand{\hc}{\mathrm{h.c.}}

\newcommand{\mAp}{m_{A^\prime}}

\makeatletter
\def\@maketitle{%
  \newpage
  \null
  \rightline{FERMILAB-PUB-21-252-T}
  \vskip 2em%
  \begin{center}%
  \let \footnote \thanks
    {\LARGE \@title \par}%
    \vskip 1.5em%
    {\large
      \lineskip .5em%
      \begin{tabular}[t]{c}%
        \@author
      \end{tabular}\par}%
    \vskip 1em%
    {\large \@date}%
  \end{center}%
  \par
  \vskip 1.5em}
\makeatother

\newcommand\KICP{Kavli Institute for Cosmological Physics, University of Chicago, Chicago, IL USA}
\newcommand\FNAL{Fermi National Accelerator Laboratory, Batavia, IL, 60510, USA}
\newcommand\HMUD{Harvey Mudd College, 301 Platt Blvd., Claremont, CA, 91711, USA}
\begin{document}
\title{\bf Multi-track Displaced Vertices at $B$-Factories}
\author[1]{Mason Acevedo}
\author[1]{Albany Blackburn}
\author[2,3]{Nikita Blinov}
\author[1]{Brian Shuve}
\author[1]{Mavis Stone}
\affil[1]{\HMUD}
\affil[2]{\FNAL}
\affil[3]{\KICP}
\date{\today}
\maketitle
\begin{abstract}
  We propose a program at $B$-factories of inclusive, multi-track displaced vertex searches, which are expected to be low background and give excellent sensitivity to non-minimal hidden sectors. Multi-particle hidden sectors often include long-lived particles (LLPs) which result from approximate symmetries, and we classify the possible decays of GeV-scale LLPs in an effective field theory framework. Considering several LLP production modes, including dark photons and dark Higgs bosons, we study the sensitivity of LLP searches with different number of displaced vertices per event and track requirements per displaced vertex, showing that inclusive searches can have sensitivity to a large range of hidden sector models that are otherwise unconstrained by current or planned searches.
  
\end{abstract}

\section{Introduction}

The nature of dark matter (DM) remains one of the most important outstanding questions in particle physics. There has been a recent shift in attention towards DM masses below the electroweak scale due to the relative lack of constraints compared to weakly interacting massive particles (WIMPs) and the promising prospects for new experimental studies \cite{Essig:2013lka,Alexander:2016aln,Battaglieri:2017aum,Beacham:2019nyx}. For  masses below about 5 GeV, it has long been established that thermal DM models require the existence of new mediators between DM and the Standard Model (SM) \cite{Lee:1977ua}. The reason is that the SM mediators in WIMP models, such as the $Z$ and Higgs bosons, are much heavier than the DM mass and consequently DM annihilation is sufficiently feeble that the DM abundance cannot be depleted enough to match observations. Viable GeV-scale DM models necessitate the addition of new mediators \cite{Fayet:2004bw}, such as a dark photon \cite{Okun:1982xi,Holdom:1985ag} or dark Higgs \cite{Silveira:1985rk,Veltman:1989vw,McDonald:1993ex,Burgess:2000yq}, which appear in models where DM is charged under a force or its mass originates from spontaneous symmetry breaking. All of this suggests an entire new hidden sector of particles coupled to DM:~not only does this make the physics of DM more analogous to that of the SM, but hidden-sector dynamics can also have other consequences, including the generation of SM neutrino masses, the creation of a baryon asymmetry, or self-interactions that affect DM structure formation. 

Most searches for hidden-sector particles at intensity-frontier experiments have focused on a relatively small subset of minimal models. In most of these scenarios, there is only one new particle that interacts with the SM, and both its production from and decay to SM states is determined by a single coupling. Alternatively, this single particle can be treated as completely invisible, either because it is meta-stable or because it decays to invisible, stable DM particles. This approach allows for powerful searches for new hidden-sector particles by exploiting the very specific production and decay modes predicted by the model, and also permits the ready comparison of different experiments using a common set of benchmarks, such as those proposed by the Physics Beyond Colliders working group \cite{Beacham:2019nyx}. 

However, as soon as additional states are added to the hidden sector, the phenomenology can change dramatically (see, \emph{e.g.},~\cite{Strassler:2006im,Alexander:2016aln}). For example, if we consider the case of a dark photon that is kinetically mixed with the SM photon, there are strong constraints on the model if the dark photon decays \emph{either} exclusively through its coupling to the SM \cite{Babusci:2012cr,Lees:2014xha,Bloise:2015wfb,Batley:2015lha,Aaij:2019bvg}, \emph{or} invisibly to unseen hidden-sector states \cite{Lees:2017lec,CortinaGil:2019nuo,NA64:2019imj}. If, however, the dark photon decays to hidden-sector states that partially decay back to the SM (for example, in the case of inelastic DM \cite{TuckerSmith:2001hy,Bai:2011jg,Izaguirre:2015zva,Izaguirre:2017bqb,Jordan:2018gcd,Mohlabeng:2019vrz,Tsai:2019mtm,Duerr:2019dmv,Duerr:2020muu,Kang:2021oes}), then neither search strategy applies, and the parameter space can be relatively unconstrained. Indeed, truly model-independent constraints from electroweak precision tests \cite{Hook:2010tw} and deep inelastic scattering \cite{Kribs:2020vyk} are sufficiently weak that they do not completely exclude dark photons as an explanation for the $4\sigma$ discrepancy between theory and experiment in $(g-2)_\mu$ \cite{Aoyama:2020ynm,Abi:2021gix}. If the mediator particle decays in a partially visible manner, then we must change our search strategies:~the mediator simply can't be discovered in the usual channels!

One of the major challenges of probing multi-particle hidden sectors is the proliferation of signatures:~if we try to constrain the hidden sector with exclusive, highly targeted searches, then a small variation in the nature of the hidden sector  invalidates the searches. It is an open problem to determine the best way to probe the model space in a manner that covers many different scenarios while limiting the number of experimental searches. At the Large Hadron Collider (LHC), the standard approach is to perform relatively inclusive searches, with results presented in a simplified model framework that facilitates re-interpretation \cite{Alves:2011wf,Abdallah:2015ter,Buchmueller:2017uqu,Alimena:2019zri}. These are supplemented by targeted, exclusive searches for well-motivated signals that are otherwise limited by large backgrounds. By contrast, at $B$-factories, which offer relatively clean environments and high luminosities that are ideal for discovering hidden-sector particles with masses below 10 GeV, most hidden-sector searches exploit exclusive signatures that target only a single, specific hidden-sector model \cite{Aubert:2009af,Lees:2012ra,Liventsev:2013zz,TheBelle:2015mwa,TheBABAR:2016rlg,Echenard:2018rfe,Adachi:2019otg,BABAR:2020oaf,BelleII:2020fag}.  The $B$-factories' hermetic nature, and excellent particle identification and reconstruction capabilities, offer interesting parallels with high-energy collider experiments such as ATLAS and CMS, making $B$-factories promising sites for generic searches targeting multi-particle hidden sectors.

In this paper, we advocate for a program of inclusive searches at $B$-factories that can dramatically improve the sensitivity of existing experiments to multi-particle hidden sectors. We focus on signatures with long-lived particles (LLPs) for a number of reasons:~LLPs are ubiquitous in hidden-sector models, and backgrounds are expected to be small. Furthermore, only a limited number of these searches currently exist at $B$-factories \cite{Liventsev:2013zz,Lees:2015rxq,BABAR:2020oaf}, and because a significant amount of work has been done at the LHC to generalize and expand the scope of LLP searches \cite{Lee:2018pag,Alimena:2019zri}, much of this knowledge gained can be applied to $B$-factories as well. We propose several extensions of existing searches for LLPs at \babar\ \cite{Lees:2015rxq} and LHCb \cite{Aaij:2020ikh}, both of which are inclusive with respect to production mechanism but require exclusive reconstruction of 2-body decays:~our proposals include searching for displaced vertices (DVs) with more than two tracks or two DVs in an event. Because of these more stringent selection criteria, such searches can relax other aspects of the analysis, such as prompt objects associated with the specific LLP production mechanism, or the exclusive reconstruction of the LLP mass; this allows a significant broadening of the applicability of the searches. We find that the projected sensitivities of our proposed searches can surpass the sensitivities for minimal dark photons or dark Higgs bosons. Our work also complements recent proposals for new, model-specific LLP searches at $B$-factories \cite{Dib:2019tuj,Filimonova:2019tuy,Duerr:2019dmv,Duerr:2020muu,Dey:2020juy}.

To motivate our approach to inclusive searches, we use SM hadronic physics as an analogy for the types of physics expected in the hidden sector \cite{Strassler:2006im}. Hadrons are  produced copiously in pairs via the strong or electromagnetic interactions, but some can only decay through the weak force due to an approximate flavour symmetry, leading to a macroscopic lifetime due to a small ratio between the LLP and $W$ masses. In the case of SM LLPs, the decay of the LLP is well-described by an effective field theory (EFT), namely the Fermi theory \cite{Fermi:1934hr}, while LLP production occurs through a low-mass mediator such as a photon, gluon, or pion.

We propose an analogous EFT approach for the study of hidden-sector LLPs. In particular, we explore several pair-production mechanisms for new LLPs, including dark photons and dark Higgs bosons. The decay of the LLP is in turn dictated by an EFT operator that couples a single LLP to SM states. This approach has several advantages:~it systematically captures all hidden-sector scenarios where the LLP decays through a heavy, off-shell mediator back to SM states; the decay width of the lightest hidden-sector state is suppressed by powers of the ratio of its mass to the EFT cutoff scale, naturally leading to long lifetimes; and it generates a rich array of phenomenological signatures. We study a number of production and EFT decay modes to show that our minimal set of inclusive LLP searches can cover a wide array of possible signals. Although our EFT framework technically only applies to off-shell degrees-of-freedom, it also characterizes the gauge-invariant SM final states that can be produced from on-shell intermediate states.

Our paper is organized as follows:~in Sec.~\ref{sec:theory_LLP}, we introduce our EFT framework for hidden-sector particle decays and argue that LLPs generically arise in such a scenario. We summarize our key results in Sec.~\ref{sec:summary}, providing details of the particular production and decay modes in Sec.~\ref{sec:production_decay} and simulation details and quantitative results in Sec.~\ref{sec:analysis}. Although our results are based on the assumption of approximately background-free signal regions, we consider the leading backgrounds and mitigation strategies in Sec.~\ref{sec:backgrounds}, and our concluding discussion is found in Sec.~\ref{sec:conclusions}.

\section{Long-Lived Particles from a Hidden Sector} \label{sec:theory_LLP}
We consider the set of models where the hidden sector is composed of a mediator, $\phi$, and an LLP, $\chi$; $\phi$ and $\chi$ could 
have any spin consistent with the decay $\phi \to \bar\chi \chi$.
We assume that both 
have masses $\lesssim10$ GeV so that they are 
accessible for on-shell production at $B$-factory energies, $e^+e^-\rightarrow \phi+X,\,\phi\rightarrow\overline\chi\chi$, where $X$ is some unspecified SM states that could be produced in association with $\phi$. Apart from the mediator, $\phi$, the only other interactions between the SM and the hidden sector occur at a much higher energy scale, $\Lambda$, which we 
take to be the scale of  new physics that mediates $\chi$ decays.

A good way to parametrize LLP decays is through effective operators~\cite{Cui:2014twa,Buchmueller:2017uqu}.\footnote{Recent comprehensive studies has also been done of EFT approaches to hidden-sector particle production \cite{Contino:2020tix,Arina:2021nqi}.} 
The reason is that the combination of mass ratios and couplings in LLP decays are typically tiny, so as long as the particles mediating the decay are at all heavier than the LLP, the EFT approach is reasonable. 
The operators mediating LLP decay are of then of the form $\chi \mathcal{O}_{\rm SM}$; $\mathcal{O}_{\rm SM}$ must be a singlet under $U(1)_{\rm EM}$ and $SU(3)_{C}$, and it is constrained by the spin of the LLP $\chi$, but is otherwise undetermined.\footnote{We will actually consider 
  the more restrictive set of $SU(3)_{C} \times SU(2)_{L} \times U(1)_Y$-invariant operators, which ensures that 
there is no requirement for new gauge-charged particles at or below the weak scale $\sim 100\;\GeV$.}
We will argue below that this leads to the observation that $\chi$ typically decays to many  electrically charged particles in the final state.

The LLP can be of any spin, but we will exclusively focus on spin-$0$ and spin-$1/2$  LLPs because these cover
most of the phenomenologically relevant signatures.\footnote{Spin-$1$ particles can be long-lived as well, including a weakly coupled
fundamental gauge boson (such as a long-lived dark photon) and composite states like the $\rho$ mesons of a confining hidden sector \cite{Berlin:2018tvf},  but these typically 
couple to conserved currents leading to two-body decays. One exception 
to this is a dark photon lighter than $2m_e$, which decays to $3$ SM photons \cite{McDermott:2017qcg}.} 
This  set-up enables a quasi-systematic way of outlining the possible LLP decay modes. We simply list the possible 
operators $\OSM$ subject to the gauge invariance constraints and organize them by operator dimension. The 
latter is key in determining the resulting LLP lifetime $\tau_\chi\equiv\Gamma_\chi^{-1}$,
\beq
\Gamma_\chi \sim m_{\chi}\left(\frac{m_{\chi}}{\Lambda}\right)^{2(n-4)},
\eeq
where $n = \dim \chi\OSM$, $\Lambda$ is the scale suppressing the higher-dimensional operator in the effective action, 
and we dropped the phase space factor which depends on the precise final-state multiplicity.
If we consider a width of $\Gamma\sim10^{-16}$ GeV (corresponding to a proper decay length of about 1 m, which is the characteristic length 
scale of the $B$-factory detectors) and $m_\chi\sim1$ GeV, then we have
\beq
\Lambda \sim 10^{16/(2n-8)}\,\,\mathrm{GeV}.
\eeq
For specific operator dimensions, we find:~for $n=5$, $\Lambda\sim10^8$ GeV; for $n=6$, $\Lambda\sim10^4$
GeV; for $n=7$, $\Lambda\sim500$ GeV; for $n=8$, $\Lambda\sim100$ GeV. Thus,
for $n\leq6$, it is very easy to get lifetimes spanning essentially the whole
range from 1 mm to $\gg1$ m while ensuring the EFT cutoffs  are above existing or prospective LHC
limits. For $n=7$, it is straightforward to get lifetimes $\sim1$ m, but
shorter lifetimes may come into conflict with LHC measurements (particularly if
the off-shell particles at scale $\sim \Lambda$ have quantum chromodynamics, or QCD,  charge). Finally, 
$n\geq8$ always gives LLPs but the lifetime may be too long to detect at $B$-factories while simultaneously satisfying
constraints from high-energy colliders. There is still a small
 acceptance for LLP decays inside the detector even for $c\tau\gg\mathrm{m}$; in this case, missing-momentum searches also constrain the model,
and these may or may not have better sensitivity than a displaced vertex search.

\begin{table}
  \centering
  \begin{tabular}{|c|c|c|c|c|}
    \hline
    $\dim \chi \OSM$ & leptonic & semi-leptonic & hadronic & photonic \\
    \hline
    5 & $\bar \ell_i \ell_i$ & & $\bar q_i q_i$, $gg$ & $\gamma \gamma$ \\
    \hline
    6 & $\nu\nu$ & & & \\
    \hline
    7 &\makecell{$\bar \ell_i \ell_j \gamma$, $\bar \ell_i \ell_j \bar\ell_k \ell_l$,\\ $\bar \ell_i \ell_j \bar\nu_k \nu_l$} & 
    \makecell{$\bar \ell_i \ell_j \bar q_k q_l$, $\bar\nu\ell \bar d u$, \\ $\color{red} \ell_i u_j u_k d_l$}
    & \makecell{$ggg$, $\bar q_i q_j \gamma$, $\bar q_i q_j g$,\\ $\bar q_i q_j \bar q_k q_l$, $\bar \nu_i \nu_j \bar q_k q_l$,\\
      $\color{red} \nu_i u_j d_k d_l$} & $\gamma\gamma$ \\
      \hline
      8 &$\color{red} \bar\ell_i \ell_j \nu_k \nu_l$ &$\color{red} \bar d_i u_j \nu_k \ell_l,\; \bar\ell_i d_j d_k d_l$ 
      & $\color{red} \bar\nu_i u_j d_k d_l,\; \bar q_i q_j \nu_k \nu_l$ & $\color{red} \nu_i\nu_j \gamma $\\
      \hline
  \end{tabular}
  \caption{Summary of final states generated in SMEFT for spin-0 LLP decay for different decay operator dimensions. 
  Final states highlighted in red violate global SM symmetries $B$ or $B-L$.
  \label{tab:s0_final_state_summary}
  }
\end{table}

\begin{table}
  \centering
  \begin{tabular}{|c|c|c|c|c|}
    \hline
    $\dim \chi \OSM$ & leptonic & semi-leptonic & hadronic & photonic \\
    \hline
    6 & $\bar \ell_i \ell_j \nu_k$ & $\ell_i u_j \bar d_k$ & \makecell{$\bar d_i d_j \nu_k$, $\bar u_i u_j \nu_k$,\\ $\color{red} u_i d_j d_k$} & $\nu\gamma$ \\
    \hline
    7 & $\bar \ell_i \ell_j \nu_k$, $\nu_i \bar\nu_j \nu_k$ & $\ell_i u_j \bar d_k$ & \makecell{$\bar d_i d_j \nu_k$, $\bar u_i u_j \nu_k$,\\ $\color{red} u_i d_j d_k$} & \\
    \hline
  \end{tabular}
  \caption{Summary of final states generated in $\nu$SMEFT for spin-1/2 LLP decay for different operator dimensions. 
  Final states highlighted in red violate global SM symmetries $B$ or $B-L$. Dimension 7 operators 
  with covariant derivatives can also generate final states with an additional 
  gauge boson $\gamma$ or $g$.
  \label{tab:s12_final_state_summary}
  }
\end{table}

The task is now straightforward:~list all SM gauge singlet operators $\OSM$ with 
\beq
\dim\OSM \leq 8 - \dim \chi, 
\label{eq:osm_dimension}
\eeq
where $\dim \chi = 1,\; 3/2$, and then classify the resulting $\chi$ products. We neglect operators involving electroweak bosons since
these cannot be produced on-shell in $B$-factories, and the additional $(m_\chi/m_W)^4$ contribution to the $\chi$ decay rate from such
operators
pushes the lifetime to be well beyond that accessible at $B$-factories. 
Since $\OSM$ is a gauge singlet, one can find all independent 
choices of this operator by harnessing the results of SMEFT (the collection of higher-dimensional operators 
constructed from SM fields alone)~\cite{Grzadkowski:2010es,Alonso:2014zka,Lehman:2014jma,Passarino:2016pzb} if the LLP is a scalar, 
or the SM + sterile neutrino EFT ($\nu$SMEFT)~\cite{delAguila:2008ir,Bhattacharya:2015vja,Liao:2016qyd} if the LLP is a fermion. The corresponding 
operators are summarized in Appendix~\ref{sec:llpeft}. In Tables~\ref{tab:s0_final_state_summary} and~\ref{tab:s12_final_state_summary}, 
we collect the parton-level final states from $\chi$ decay enabled by these operators 
for spin-$0$ and spin-$1/2$ LLPs, respectively. 

One pronounced feature of  
final states from the EFT framework is that the majority  feature multiple charged particles. 
Note that when the operator involves quarks, the actual decay final states are generated by hadronization which 
may lead to a different number of charged particles than na\"ively expected depending on the LLP mass.  
We use the above observation to develop a generic set of analysis proposals for multi-track displaced
vertex searches that can probe 
many of these final states.

\section{Summary of Main Results} \label{sec:summary}
Before turning to our detailed results, we provide a brief summary of our primary conclusions. We propose inclusive displaced vertex searches, studying the signal acceptance and efficiency as a function of the number of displaced vertices per event, $\Ndv$, and the number of tracks required per vertex, $\Ntr$. We expect that searches with $\Ndv\ge2$ and $\Ntr\ge3$ should be background free, although searches with lower track and/or vertex multiplicities are possible especially if additional selections (like vertex track mass cuts) are implemented. Our findings include:
\begin{itemize}

  \item \emph{A minimal set of searches has broad coverage of different production modes, decay modes, and LLP lifetimes.} While the detailed acceptances depend on model specifics, all of the simplified production and decay modes we consider have appreciable acceptance for LLP searches with different $\Ndv$ and $\Ntr$, including $\Ndv\ge2$ and $\Ntr\ge3$. In most cases, the sensitivity is sufficient to probe interesting parameter space motivated by $(g-2)_\mu$ or dark matter models.\footnote{While the $(g-2)_\mu$ anomaly can be explained by new particles, recent lattice calculations of hadronic vacuum polarizations in the SM alone can also bring the theoretical 
    prediction into agreement with the BNL and FNAL results~\cite{Borsanyi:2020mff}.} 
    Expected coupling sensitivities are competitive with or surpass existing searches for fully visible or invisible dark photons and dark Higgs bosons.

\item \emph{Signal acceptance falls with stricter requirements on $\Ndv$ and/or $\Ntr$, but not as severely as we expect backgrounds to be suppressed.} For combinatoric and other rare backgrounds, increasing the track multiplicity, vertex multiplicity, track mass, etc., can effectively suppress the backgrounds. Meanwhile, the signal acceptance is always somewhat reduced by the tighter selections, but usually the residual acceptance is still sufficient to probe interesting parameter space.

\item \emph{LLPs that decay to partially or fully hadronic final states can give large multiplicities of charged tracks in the final state.} Even when the quark-level operators suggest a low multiplicity of charged final states, hadronization effects (whether in a parton shower or chiral perturbation theory) can give higher multiplicities of charged pions or kaons in the final state. This allows for a significant signal acceptance even with tighter cuts $\Ntr\ge3$. Heavier LLPs more readily pass selections requiring more than two tracks per vertex due to more pronounced effects of parton showering or heavy intermediate QCD resonances, while lighter LLPs can suffer a severe acceptance penalty when requiring more than two tracks per vertex for hadronic decays.

\item \emph{Hadronization uncertainties can be significant depending on the model, and are greatest for low-mass LLPs.} When LLP decay operators resemble those in semileptonic  decays of SM particles, the effects of hadronization are relatively well-understood. For other operators, the uncertainties are larger, especially for LLP masses such that the typical hadronic 4-momentum invariant is below 1 GeV. Even when different models of hadronization disagree, we still find a reasonable signal acceptance for all hadronization models that allows DV searches to probe interesting parameter space. Crucially, the search itself can be done without worrying about theory uncertainties, as it is only the mapping of the results of a search to a particular model that is sensitive to such uncertainties. For more details, see Appendix \ref{sec:hadronization}.

\item \emph{Exclusive signal reconstruction  would reduce or eliminate the sensitivity to many of the models we consider. However, some model-dependent selections could help suppress backgrounds if needed.} For several of our models, it is challenging to fully reconstruct the LLP decay and/or production signal, either due to invisible particles or the fact that there are many possible LLP decay modes. For examples, many modes include $\pi^0\rightarrow\gamma\gamma$ in the LLP decay chain that could be difficult to associate with the displaced vertex. These models are not covered by the existing model-independent search by \babar\, which requires a reconstruction of the LLP mass in displaced track pairs \cite{Lees:2015rxq}. However, if a totally inclusive search is limited by backgrounds, it is possible to improve sensitivity by applying a few additional selections that depend on the model. These could include particle identification requirements on one or more of the tracks at the displaced vertex, tagging a  photon from initial-state radiation (ISR) associated with dark photon production in the radiative return process, and so on. We encourage experimentalists to apply as few of these requirements as is feasible to ensure the broadest coverage of any given search.

\end{itemize}

\section{Production and Decay of Long-Lived Particles} \label{sec:production_decay}
The central observation of the previous sections is that many operators 
connecting the hidden sector to the SM induce decays to several charged particles, 
giving rise to multi-track events. Thus, it is useful to develop a widely applicable 
search strategy to target these generic final states. In this section, we identify 
several test cases that lead to events with multiple tracks emanating from displaced vertices and that can be probed with this approach. 
As a reminder we denote the mediator by $\phi$ and the LLP by $\chi$, although these can each be scalars, fermions, or vectors depending on the interaction.
We consider leptonic, semi-leptonic, and hadronic final states. While these final states arise 
from a multitude of operators, for simplicity of simulation we focus on the following 
operators for concreteness:
\begin{itemize}
  \item leptonic final states (scalar LLP)
    \beq
    \chi\bar \mu \mu \bar e e;
    \eeq
  \item fully and non-fully reconstructible semi-leptonic final states (fermion or scalar LLP) 
    \beq
    \left(\bar u_{R}\gamma_\mu d_{R} \right)\left( \bar\ell_{R} \gamma^\mu \chi\right),\;\; \chi (\bar{\nu}_L e_R)(\bar{d}_L u_R)  ;
    \eeq
  \item hadronic final states (scalar LLP)
    \beq
    \chi (\bar{u}_R \gamma^\mu u_R)( \bar{u}_R \gamma_\mu u_R).
    \eeq
\end{itemize}
We note that the specific flavour combinations chosen are somewhat arbitrary; we selected these either to allow for the exploration of the widest kinematic 
range in LLP decays by focusing on light flavours, or because some hadronic decays can be modelled by analogy with certain SM 
processes.
We implement these models in the \texttt{Universal FeynRules Output (UFO)} format~\cite{Degrande:2011ua} using \texttt{FeynRules}~\cite{Alloul:2013bka}.
These models are then used in \madgraph~\cite{Alwall:2011uj,Alwall:2014hca} to simulate the relevant processes.
The operators containing quarks require special attention since they must be hadronized to generate physical 
final states. We describe our treatment in detail in Appendix~\ref{sec:hadronization}.

We must also specify a production mechanism for the LLPs, which can be parametrized by a few benchmark simplified models. For simplicity, we study two production modes:
\begin{enumerate}
  \item {\bf Photon associated production,} namely $e^+e^-\rightarrow \gamma\phi$, $\phi\rightarrow\chi\bar\chi$. This occurs for vector-portal models through the radiative return process (either through kinetic mixing with the SM photon or direct coupling of the vector to electrons)~\cite{Lees:2014xha,Lees:2017lec}, or scalars with direct couplings to electrons.  It is also valid for axion-like particles that possess a $\phi\gamma\gamma$ coupling (and the production proceeds through $e^+e^-\rightarrow \gamma^*\rightarrow \gamma \phi$), although the kinematics is distinct from radiative return. A common feature to both scenarios is the monochromatic photon that can, in principle, be used to discriminate against backgrounds with a bump hunt in the photon energy, provided it is tagged. The multiplicity of tracks and photons from the production process is low, which could allow us for discrimination against QCD processes. For concreteness, we use a dark photon model to implement this production mechanism (see Sec.~\ref{sec:rad_ret} for more details).

  \item {\bf Heavy-flavour production.} Mediators lighter than the $B$ meson can be produced in flavour-changing neutral current decays such as $B\rightarrow K\phi$, $\phi\rightarrow\chi\bar\chi$. If necessary, it is possible to suppress continuum QCD backgrounds using event-shape variables and by requiring that the event (with $B\rightarrow K\phi$ candidate) matches the constrained kinematics of $\Upsilon(4S)\rightarrow B\bar{B}$. This production mode is also interesting because there should be a double displaced vertex (one ``vertex'' from the $K$ production and another from the LLP decay). These are generally high track-multiplicity events, which can  result in larger backgrounds. For concreteness, we use a dark Higgs model to implement this production mechanism (see Sec.~\ref{sec:B_dec} for more details).

\end{enumerate}
In both cases, we assume 100\% decay of the mediator into LLPs. These production modes are  well-motivated and   also encompass a range of mediator and LLP kinematics.  

We emphasize that our proposed searches are designed to be inclusive of the production mechanism. As a result, sufficiently inclusive searches can actually be sensitive to broader classes of scenarios, such as:
\begin{itemize}

\item \underline{Heavy-flavour leptonic forces:}~If the mediator is an $L_\mu-L_\tau$ gauge boson \cite{He:1990pn,He:1991qd} or a leptophilic scalar \cite{Chen:2015vqy,Batell:2016ove}, the dominant production of the LLPs will be in association with $\mu^+\mu^-$ or $\tau^+\tau^-$ pairs. This channel was recently studied for dielectron LLP decays in the \babar~leptophilic dark scalar search~\cite{BABAR:2020oaf}.

    \item \underline{Dark-flavour non-diagonal production}:~As an example, consider inelastic dark matter coupled to a dark photon, $e^+e^-\rightarrow \gamma A'$, $A'\rightarrow\chi_2\chi_1$, where $\chi_1$ is a stable invisible particle and $\chi_2$ is the LLP \cite{Izaguirre:2015yja,Izaguirre:2015zva,Izaguirre:2017bqb,Jordan:2018gcd,Mohlabeng:2019vrz,Tsai:2019mtm,Duerr:2019dmv,Duerr:2020muu,Kang:2021oes}. This example is an illustrative model that predicts only a \emph{single} LLP, showing the importance of single-LLP searches ($\Ndv=1$) where possible. In this minimal model, the $\chi_2$ exclusively decays through the same dark photon coupling that allows its production, but more general models could additionally include different decays of the $\chi_2$.

\item \underline{Neutrino production}:~ If a neutral fermion $N$ mixes with the SM neutrino, then $N$ is produced in meson decays such as $B\rightarrow \ell N+X$ \cite{Gorbunov:2007ak} (where $X$ usually includes a $D$ meson, but could also be ``nothing'', in which case we have exclusive production in leptonic decays). If $N$ decays through the same neutrino mixing,  then we  recover the Majorana neutrino scenario that is the subject of a search at Belle \cite{Liventsev:2013zz}, although that search only examined a single exclusive decay mode for the LLP, $N\rightarrow\ell^\pm\pi^\mp$.  If the model additionally contains a singlet scalar $S$, the dominant decay could instead be $N\rightarrow S\chi$, where either $S$ or $\chi$ could be invisible stable particles or LLPs.

\item \underline{Strongly interacting hidden sectors}:~An example of this is a confining hidden valley \cite{Strassler:2006im} where a dark photon decays into hidden-sector quarks that then shower and hadronize \cite{Hochberg:2015vrg}. This could encompass Strongly Interacting Massive Particles (SIMPs) \cite{Hochberg:2014dra,Hochberg:2014kqa} (for instance, the $A'\rightarrow \rho_{\rm D}\pi_{\rm D}$ signature \cite{Berlin:2018tvf}), either with $m_{A'}\sim \Lambda_{\rm D}$ or $m_{A'}\gg \Lambda_{\rm D}$ where $\Lambda_{\rm D}$ is the dark-sector confinement scale.

\end{itemize}
In several of the above examples, the dark sector goes beyond the ``next-to-minimal'' set-up we consider (a mediator, LLP and a higher dimensional decay operator), and although each merits its own
careful study, it is evident that inclusive LLP searches at $B$-factories should encompass a variety of signals in terms of number of charged particles per LLP decay and vertex multiplicity.

\subsection{LLP Production:~Radiative Return (Dark Photon)}
\label{sec:rad_ret}
The main mechanism by a which an electron-coupled vector or scalar mediator, such as a dark photon, 
is produced at $B$-factories is via radiative return. In the radiative-return process, ISR photons drive the $e^+e^-$ center-of-mass (CM) energy to resonance. We will use the dark photon, $A'$,  with the kinetic mixing 
interaction $(\varepsilon/2) F^{\mu\nu} F_{\mu\nu}^\prime$
as the specific mediator example for our implementation of radiative return. The corresponding 
Lagrangian that enables $A'$ production and decay into LLPs is
\beq
\mathcal{L} \supset \varepsilon e A^\prime_\mu J^\mu_{\mathrm{EM}} + g_D A^\prime_\mu J^\mu_{\chi},
\label{eq:dark_photon_interaction}
\eeq
where $J^\mu_{\mathrm{EM}}$ ($ J^\mu_{\chi}$) is the electromagnetic (dark) current, 
and $g_D$ is the coupling between  $A'$ and the LLP.

\begin{figure}
  \centering
  \includegraphics[width=0.47\textwidth]{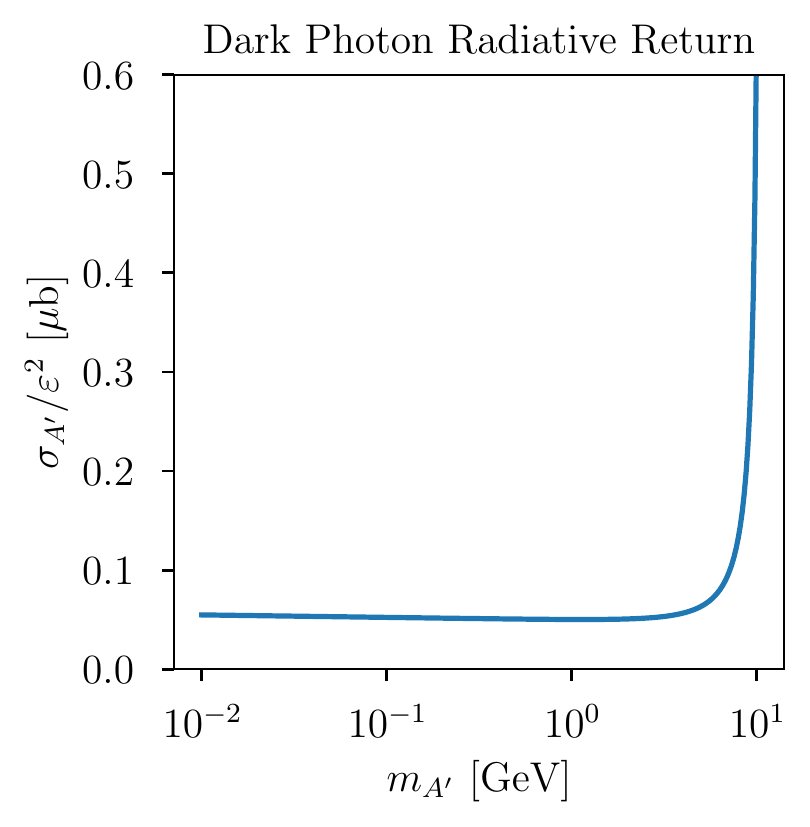}
  \caption{Radiative return cross section in $\mu\mathrm{b}$, divided by squared kinetic mixing, $\varepsilon^2$, for on-shell dark photon production.\label{fig:rad_ret_xsec}}
\end{figure}

The total radiative return cross section is, in the narrow-width approximation  
(see, \emph{e.g.,} Ref.~\cite{Karliner:2015tga}),
\beq
\sigma_{A'}(s) = \left[\frac{4\pi^2 \alpha \varepsilon^2}{s}\beta_f \left(\frac{3-\beta_f^2}{2}\right)\right]
\int_{m_{A\prime}^2/s}^{s} \frac{dx}{x} f_e(x) f_e\left(\frac{m_{A'}^2}{xs}\right),
\label{eq:rr_total_cross_section}
\eeq
 where $\beta_f^2 = 1 - 4m_e^2/m_{A'}^2$. 
In the above expression, the first factor is the resonant 
annihilation cross section for $e^+ e^-$ (divided by the Mandelstam variable $s$ and with the delta function stripped off), 
while the second factor is the effective $e^+ e^-$ parton luminosity with $f_e$ being the 
electron distribution function which encodes the probability of finding an electron or positron with momentum fraction $x$. 
Note that Eq.~\eqref{eq:rr_total_cross_section} includes the part of phase space in which the 
photon is collinear with the beams and therefore unobservable; this cross section can be a factor of $\ln (s/m_e^2) \sim 20$ 
larger than if a photon is required in the detector acceptance (a requirement used in existing dark photon searches at \babar~\cite{Lees:2014xha,Lees:2017lec}). 
We show the radiative return cross section as a function of $A'$ mass in Fig.~\ref{fig:rad_ret_xsec}.

To simulate events with radiative return production of $A'$, we sample $x$, the lepton momentum fraction, 
from the parton luminosity function using the Kuraev-Fadin distribution~\cite{Kuraev:1985hb} from Ref.~\cite{Nicrosini:1986sm}. 
The kinematics of $A'$ and its decay products in the lab frame can then be reconstructed by noting that $A'$ is 
produced nearly at rest in the CM frame in the narrow-width approximation. We can then boost 
the combined decay chain of $A'$ to LLPs and LLPs to SM (generated using \madgraph) into the lab frame. While a publicly-available lepton ISR plugin exists that would have allowed us to include ISR directly in \madgraph~\cite{Chen:2017ipx,Li:2018qnh}, we 
found that it had difficulty populating the phase space around the narrow $A'$ resonance. 
We have verified that our procedure gives the correct parton luminosity, proportional to the integral piece 
in Eq.~\eqref{eq:rr_total_cross_section}, by matching to the results of Ref.~\cite{Karliner:2015tga,Greco:2016izi}. 
We also compared the results of Eq.~\eqref{eq:rr_total_cross_section} with \madgraph including ISR, finding that the cross sections match
at those parameter points for which \madgraph events were successfully generated.   

For $m_{A'} < \sqrt{s}$ the expected number of LLP pairs produced for the full Belle II integrated luminosity is 
\beq
N_{\chi\bar\chi} \sim 10^{9} \left(\frac{\varepsilon}{0.03}\right)^2,
\label{eq:rad_ret_evt_sample}
\eeq
where we normalized the rate to the kinetic mixing that saturates model-independent bounds 
from precision electroweak observables~\cite{Hook:2010tw}; for $\mAp \approx 2\;\GeV$, $\varepsilon \approx 0.03$ 
also resolves $(g-2)_\mu$. While for minimal assumptions (purely visible or invisible decays of $A'$) 
these ``large'' kinetic mixings are excluded by exclusive searches~\cite{Lees:2014xha,Lees:2017lec}, 
this parameter space can be open if the $A'$ decays do not 
satisfy the selection criteria in those analyses. 

\begin{figure}
  \centering
  \includegraphics[width=0.47\textwidth]{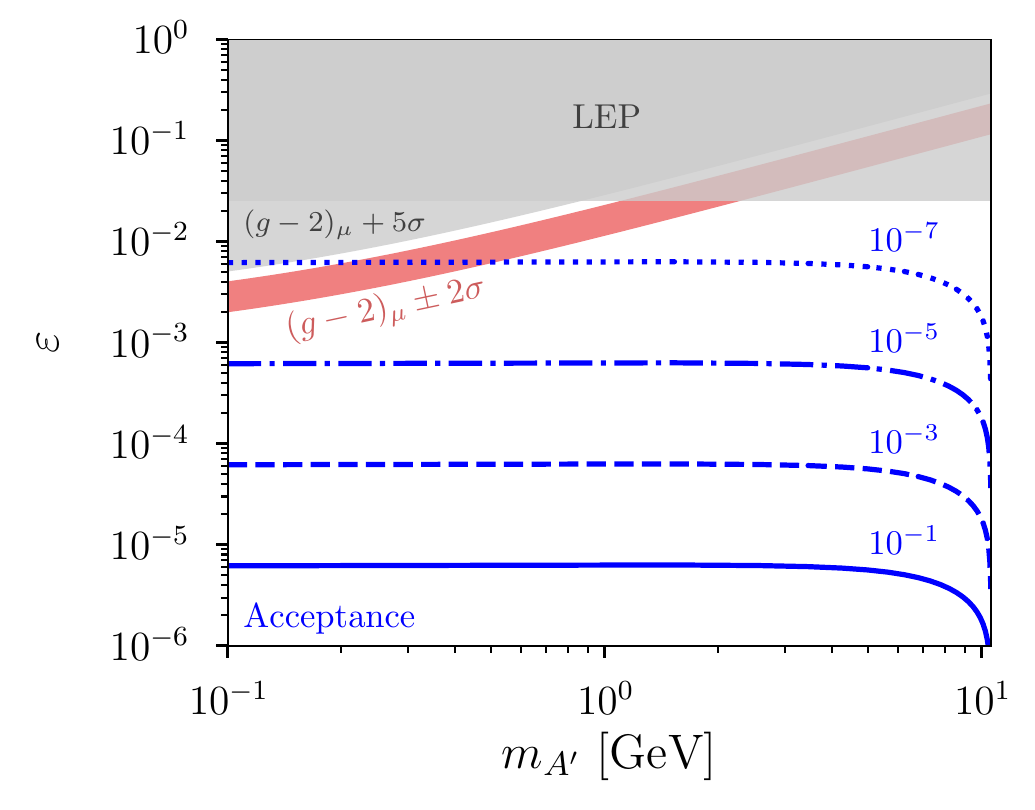}
  \caption{Sensitivity to the dark photon mass and kinetic mixing, $\varepsilon$, under the assumption that selections with the indicated signal acceptance are background-free and can be discovered with 10 signal events after cuts with $\mathcal{L}=50\;\ab^{-1}$ at Belle II. We also show model-independent limits on the dark photon
  parameters~\cite{Pospelov:2008zw,Hook:2010tw}, as well as the band favoured by the $(g-2)_\mu$ anomaly~\cite{Pospelov:2008zw}.
  \label{fig:gm2_darkphoton}
  }
\end{figure}

The large number of allowed LLP events in Eq.~\eqref{eq:rad_ret_evt_sample} then suggests 
that novel analyses can  offer significant discovery potential (including for $(g-2)_\mu$-preferred parameter space) even with small signal acceptance. To illustrate this, we show in Fig.~\ref{fig:gm2_darkphoton} the schematic background-free discovery potential in the dark photon parameter space as a function of the signal acceptance. We require that 10 signal events pass all selections with the indicated acceptance and the full integrated luminosity of Belle II, $\mathcal{L}=50\;\ab^{-1}$. It is evident that even if searches require very stringent selections to eliminate backgrounds, leading to a very low signal acceptance ($\sim10^{-7}$), they can still be sensitive to  parameters motivated by $(g-2)_\mu$. 

The coupling of $A'$ to the LLP current, $g_D A^\prime_\mu J^\mu_{\chi}$, suggests that $\chi$ is charged under a dark $U(1)$ gauge symmetry. This gauge symmetry must be broken to allow $\chi$ to decay as an LLP back to SM particles. This can be accomplished in several ways:~for example, the $\chi\mathcal{O}_{\rm SM}/\Lambda^{n-4}$ operators allowing $\chi$ to decay could have an additional insertion of some dark Higgs vacuum expectation value, $\langle h'\rangle/\Lambda$. Alternatively, $\chi$ could mix with a heavy Majorana singlet state $\xi$ after dark symmetry breaking, and it is secretly a $\xi\mathcal{O}_{\rm SM}/\Lambda^{n-4}$ effective operator that allows $\chi$ to decay to the SM. We are agnostic about which mechanism allows $\chi$ to decay to the SM in the dark photon model, in part because we treat the $\chi$ lifetime as a free parameter and  these model-dependent considerations simply lead to adjustments in the relationship between the $\chi$ lifetime and the EFT scale, $\Lambda$. 

We finally note that the interactions of Eq.~\eqref{eq:dark_photon_interaction} imply an irreducible branching fraction of $A'$ into SM final states $f$, which scales as $\mathrm{BR}(A' \to \bar f \bar f) \sim \varepsilon^2 e^2/g_D^2$. For sufficiently large  values of $\varepsilon$, this suggests that $A'$ can be probed with the usual visible dark photon searches such as Ref.~\cite{Lees:2014xha}. Since the sensitivity of visible searches in this limit scales as $\varepsilon^4$, the visible decays are negligible when $\varepsilon\lesssim0.01$ and $g_D\sim1$ even with the expected full Belle II luminosity. We therefore focus on the parameter space where the dominant sensitivity is instead provided by the multi-track searches described in Sec.~\ref{sec:analysis}.

\subsection{LLP Production:~\texorpdfstring{$B$}{B} Meson Decays (Dark Higgs)}
\label{sec:B_dec}

We  consider a Higgs-portal coupling between a scalar mediator, $\phi$, and the LLP, $\chi$. After electroweak symmetry breaking, the SM and dark Higgs states mix with angle $\theta$, giving a tree-level coupling to quarks and LLPs:
\begin{eqnarray}
\mathcal{L} &\supset& -\theta\frac{m_q}{v}\phi\bar{q}q - y\,\phi\bar\chi\chi,
\end{eqnarray}
where the coupling $y$ of $\phi$ to LLPs is assumed to be sufficiently large such that $\mathrm{BR}(\phi\rightarrow\chi\bar\chi)\approx100\%$. In loop-induced $B$ decays, the dark Higgs benefits from the large coupling of $\phi$ to top quarks, and it therefore has a large production rate in flavour-changing decays of $b$ quarks, $b\rightarrow s\phi$. For  $m_\phi \ll m_b$, the inclusive $\phi$ production rate in $B$ meson decays is \cite{Boiarska:2019jym}
\begin{eqnarray}\label{eq:massless_inclusive}
\mathrm{BR}(B\rightarrow X_s \phi) &\approx& 3.3\,\theta^2.
\end{eqnarray}
The cross sections for $B^+B^-$ and $B^0\bar{B}^0$ production  are each 550 pb at the $\Upsilon(4S)$ resonance. Before selections, and with  $50\;\ab^{-1}$ of integrated luminosity, this gives  a number of $\chi\bar{\chi}$ pairs equal to
\begin{eqnarray}
N_{\chi\bar\chi} &\sim& 40\left(\frac{\theta}{10^{-5}}\right)^2.
\end{eqnarray}

For larger $\phi$ masses, it is more appropriate to sum over exclusive decay modes $B\rightarrow K\phi$, where we include various kinematically accessible kaon resonances. For $4.3\,\,\mathrm{GeV}<m_\phi < 4.78\,\,\mathrm{GeV}$, only a single decay into the lightest kaon occurs with a rate of about 10\% of the inclusive rate from Eq.~\eqref{eq:massless_inclusive}, while for somewhat smaller $\phi$ masses the sum of exclusive production modes is comparable to the rate predicted by the quark-level calculation. In our analysis, we use the rates from Ref.~\cite{Boiarska:2019jym}.

We have used \texttt{FeynRules} to implement a \texttt{UFO} model containing $\Upsilon(4S)$, $B^\pm$, and various $K^\pm$ mesons, along with the dark Higgs $\phi$, and we simulate $\Upsilon(4S)\rightarrow B^+B^-,B^+\rightarrow K^+\phi$, while the other $B^-$ decay mode is unspecified.\footnote{For $\theta\ll1$, the probability that both $B$ mesons decay into $\phi$ is negligible.} The effective flavour-changing $\phi-B-K$ coupling is implemented at tree level, and the relative couplings of the different kaon states are chosen to match the relative exclusive decay rates from Ref.~\cite{Boiarska:2019jym}. Since we are interested in LLPs that are sufficiently heavy that they decay into multiple charged particles, we are mostly interested in the case where $m_\phi$ is not much smaller than $m_B-m_K$ and hence the production of $\phi$ is well approximated by summing over exclusive decay modes.

\section{Simulation and Analysis}
\label{sec:analysis}
In this section, we demonstrate that searches for displaced vertices containing multiple charged tracks are broadly sensitive 
to the generalized LLP decays described by EFT operators as described in the previous sections. 
We focus on two production channels, radiative return and rare $B$ meson decays, which 
we simulate as described in Sections~\ref{sec:rad_ret} and~\ref{sec:B_dec}. The resulting LLPs are 
decayed in \madgraph and hadronized in \pythia \cite{Sjostrand:2007gs}, or manually by matching to chiral perturbation theory as discussed in Appendix~\ref{sec:hadronization}.
We analyze these events in the context of the Belle II experiment at the SuperKEKB facility. 

The Belle II experiment studies $e^+e^-$ collisions at $\sqrt{s} = 10.58\;\GeV$ (with $E_{e^+} = 4\;\GeV$ and $E_{e^-} = 7\;\GeV$) using a detector described in 
detail in, \emph{e.g.,} Refs.~\cite{Abe:2010gxa,Kou:2018nap}. Over its lifetime, Belle II will collect 
$50\;\ab^{-1}$ of integrated luminosity, yielding unprecedented sensitivity to rare processes. 
For our LLP searches, the most important detector subsystems are the vertex detector (VXD), with layers starting only $1.4$ cm from the 
interaction point and going out to $17$ cm, and the central drift chamber (CDC), which extends radially out to $1.1$ m. 
Both the VXD and CDC provide polar angular coverage between $17^\circ$ and $150^\circ$ in the lab frame. 
The displaced vertex (DV) detection efficiency is transverse radius-dependent and we follow Ref.~\cite{Duerr:2019dmv} 
in taking~\footnote{These efficiencies depend on the particle track type in the vertex (\emph{e.g.,} electron versus muon), but 
we take them to be equal for simplicity.} 
\beq
\text{DV efficiency} = \begin{cases}
1 & 0.2\;\cm \leq r_\perp \leq 17\;\cm \\
  0.3 & 17\;\cm < r_\perp \leq 60\;\cm \\
  0 & r_\perp > 60 \;\cm
\end{cases}
\eeq
for $-55\;\cm < z < 140\;\cm$ and zero outside. We also account 
for the 42 mrad beam ``half-crossing'' angle~\cite{Abe:2010gxa} as described in Ref.~\cite{Duerr:2019dmv}. 
In identifying tracks, we demand charged particles to be within the geometric acceptance and to have transverse momentum $p_T > 100\;\MeV$; 
we also assign a detection efficiency per track of $0.9$. While track-finding efficiencies are strongly dependent on  $p_T$, impact parameter, and decay position, the 0.9 value we use is a simple  way of incorporating tracking inefficiencies  that gives, on average, results consistent with expectations from the Belle II detector \cite{Bertacchi:2020eez}.
Same-sign tracks with angular separation $\theta < 0.05$ are merged and counted as one, although we find this has little effect on our results.

We use these definitions to find tracks and displaced vertices in our event samples and 
study the acceptance (and the resulting reach) for different 
choices of the minimum number of DVs per event, $\Ndv$, and the 
minimum number of tracks per DV, $\Ntr$.

We estimate the sensitivity of LLP searches assuming there are no 
backgrounds, and thus a 10-event signal detection is sufficient for discovery. 
We expect that this is a good approximation 
for searches requiring $\Ndv\geq 2$ and $\Ntr \geq 3$. 
While even this stringent selection is sensitive to many LLP production and 
decay channels, it is also interesting to consider looser selections, 
which may require additional background mitigation strategies. 
We will describe the relevant backgrounds and possible additional cuts in Sec.~\ref{sec:backgrounds}, 
but it is important to note that these cuts can be used to reduce backgrounds without 
significantly impacting signal acceptance across a broad range of LLP parameter space.

\subsection{Radiative Return Production of a Dark Photon}
\begin{figure}
      \centering
      \includegraphics[width=0.47\textwidth]{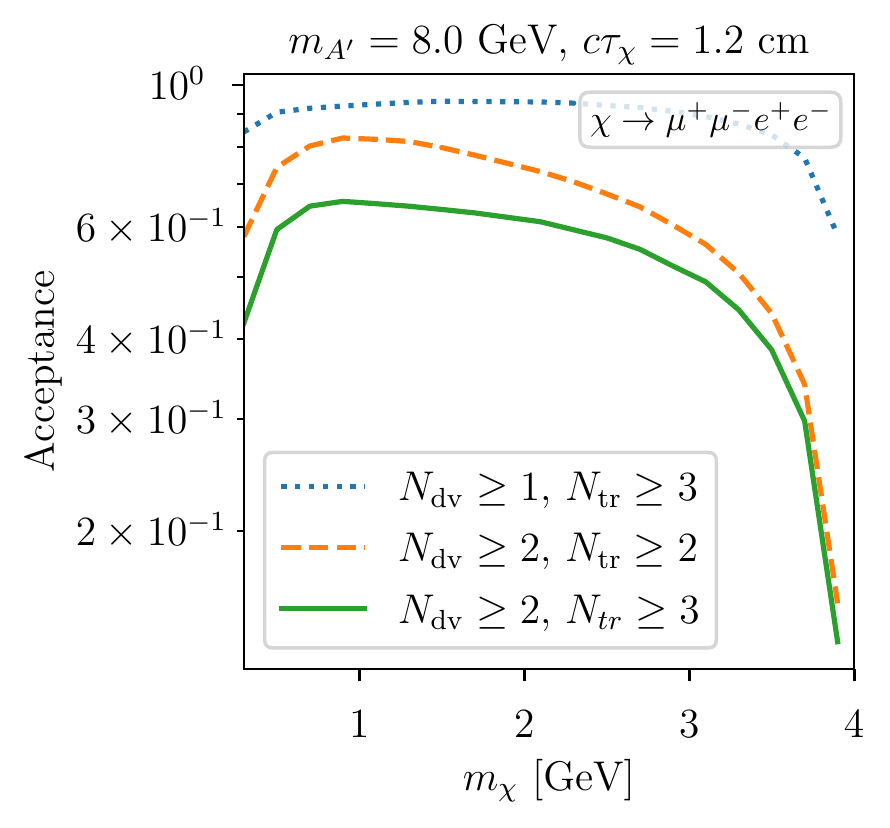}
      \includegraphics[width=0.47\textwidth]{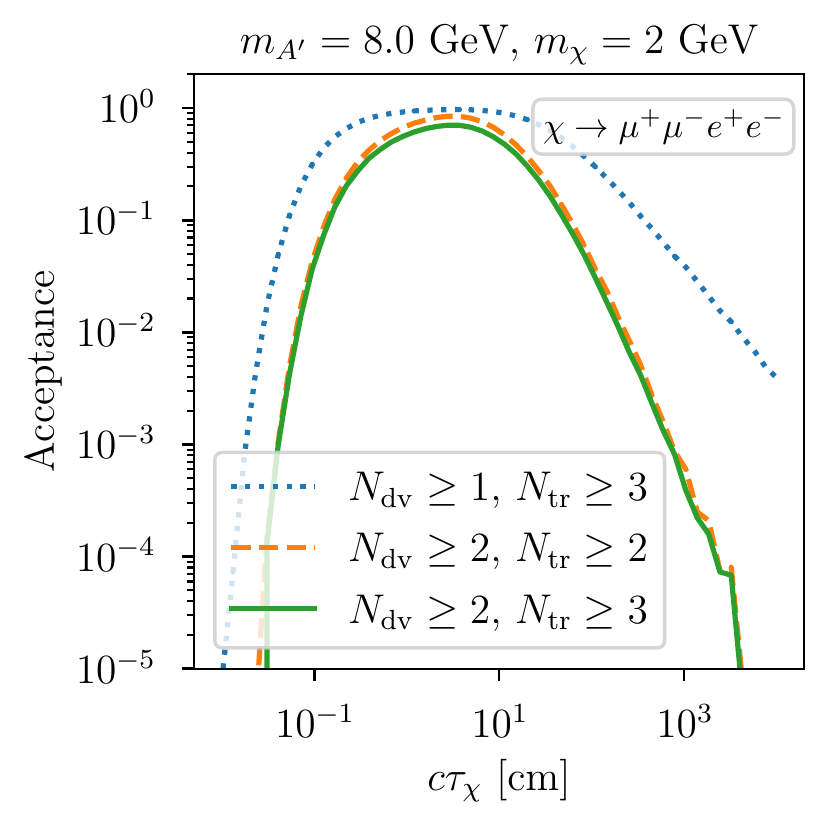}
      \caption{
        Slices of the event acceptance probability for scalar LLPs $\chi$ decaying to $e^+ e^- \mu^+ \mu^-$ for dark photon production with $m_{A'}=8\;\GeV$. The 
        left (right) panel shows the acceptance as a function of the LLP mass for $c\tau_\chi\approx 1\;\cm$ 
        (as a function of $c\tau_\chi$ for $m_\chi = 2\;\GeV$.)
      \label{fig:acceptance_slices_eemumu}}
    \end{figure}

\begin{figure}
      \centering
      \includegraphics[width=0.47\textwidth]{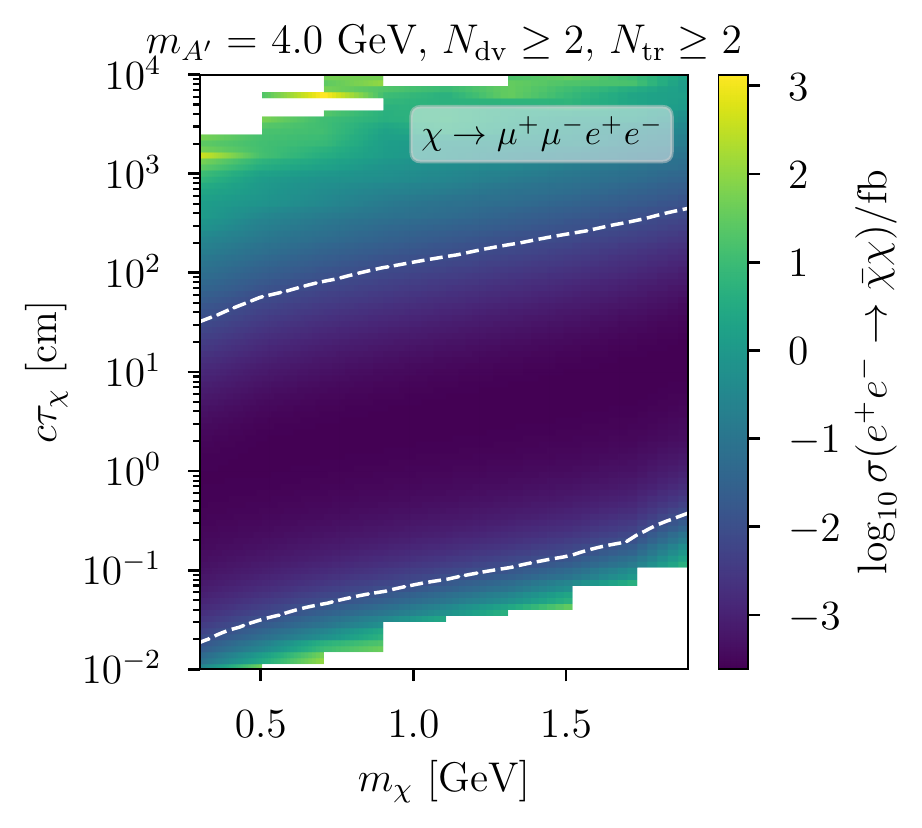}
      \includegraphics[width=0.47\textwidth]{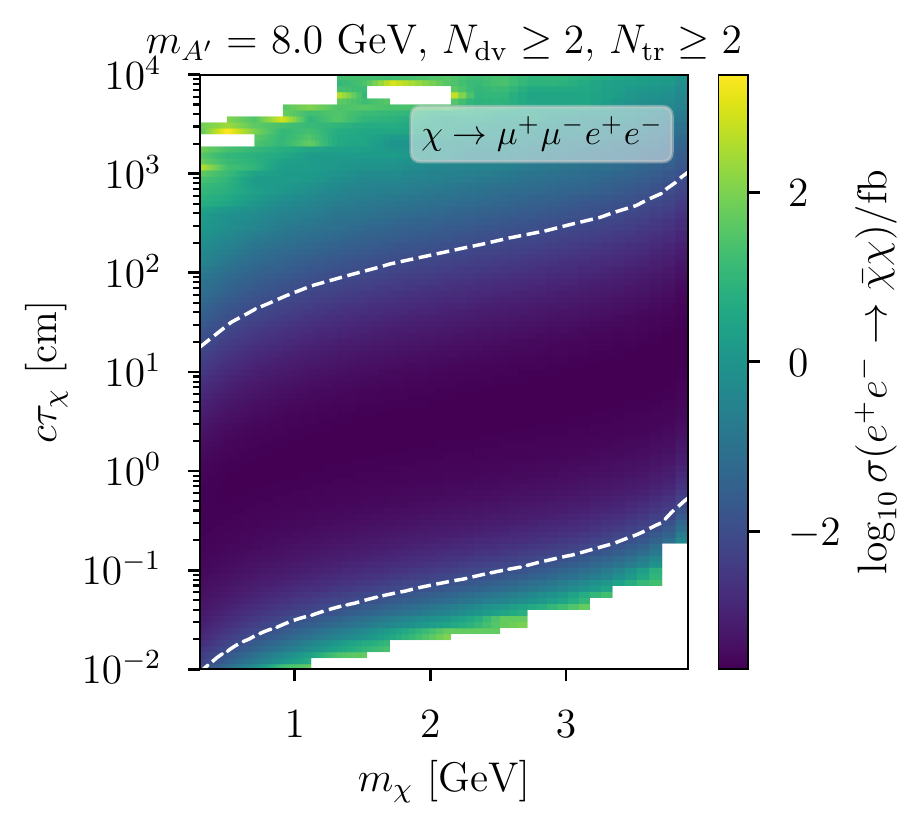} \\
      \includegraphics[width=0.47\textwidth]{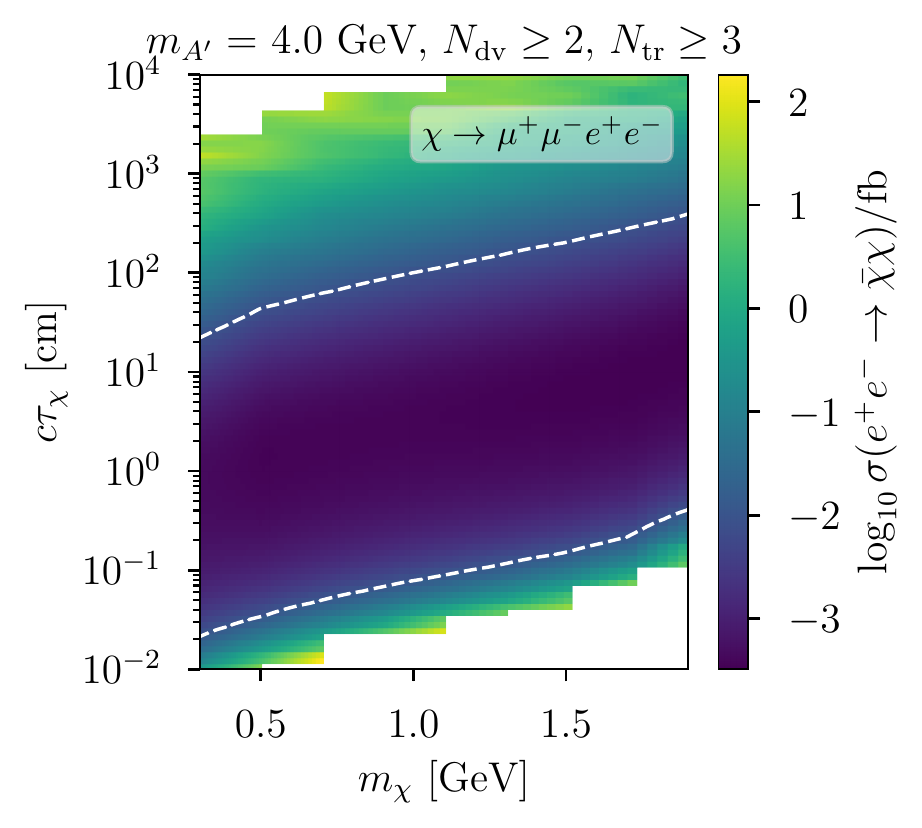}
      \includegraphics[width=0.47\textwidth]{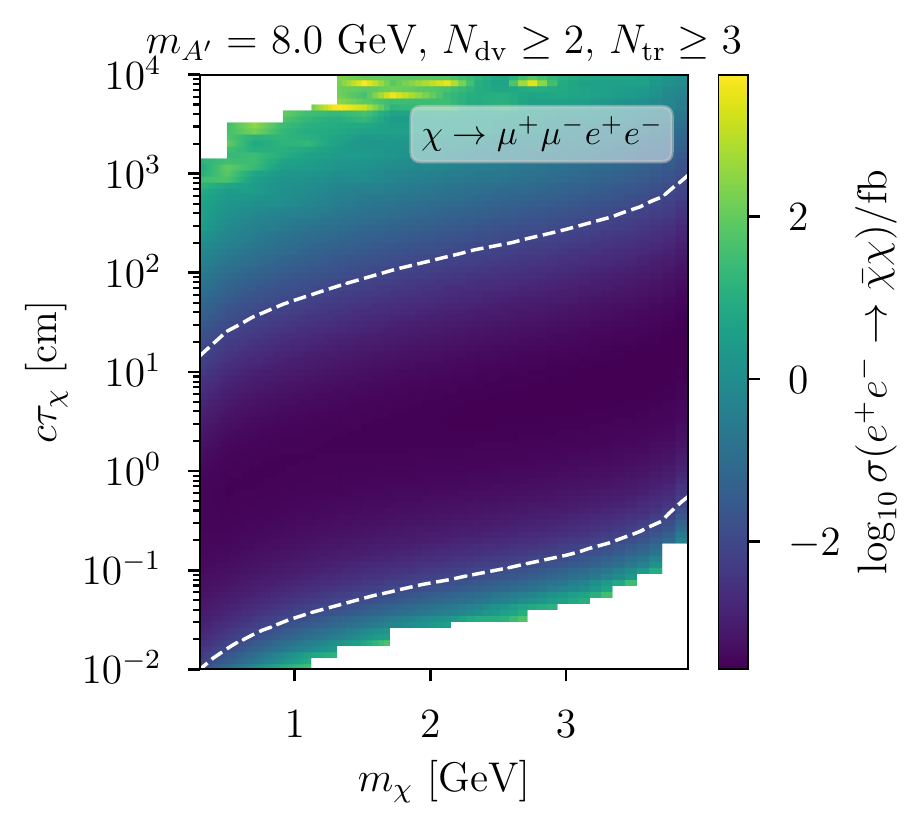} 
      \caption{Cross-section sensitivity in $\fb$ of Belle II based on 10 signal events in a $50\;\ab^{-1}$ dataset for 
        scalar LLPs $\chi$ decaying to $e^+ e^- \mu^+ \mu^-$ and the dark photon production mode. The upper (lower) row shows 
        the result of 
        demanding at least 2 (3) charged tracks from two displaced vertices. The 
        left (right) column fixes $m_{A'} = 4$ ($8$) GeV. The white dashed contours 
        correspond to a production cross section of $10^{-2}\;\fb$, which is 
        roughly equivalent to $\varepsilon \sim 10^{-5}$ for $m_{A'} \leq 8\;\GeV$.
      \label{fig:xsec_sensitivity_eemumu}}
    \end{figure}

\begin{figure}
      \centering
      \includegraphics[width=0.47\textwidth]{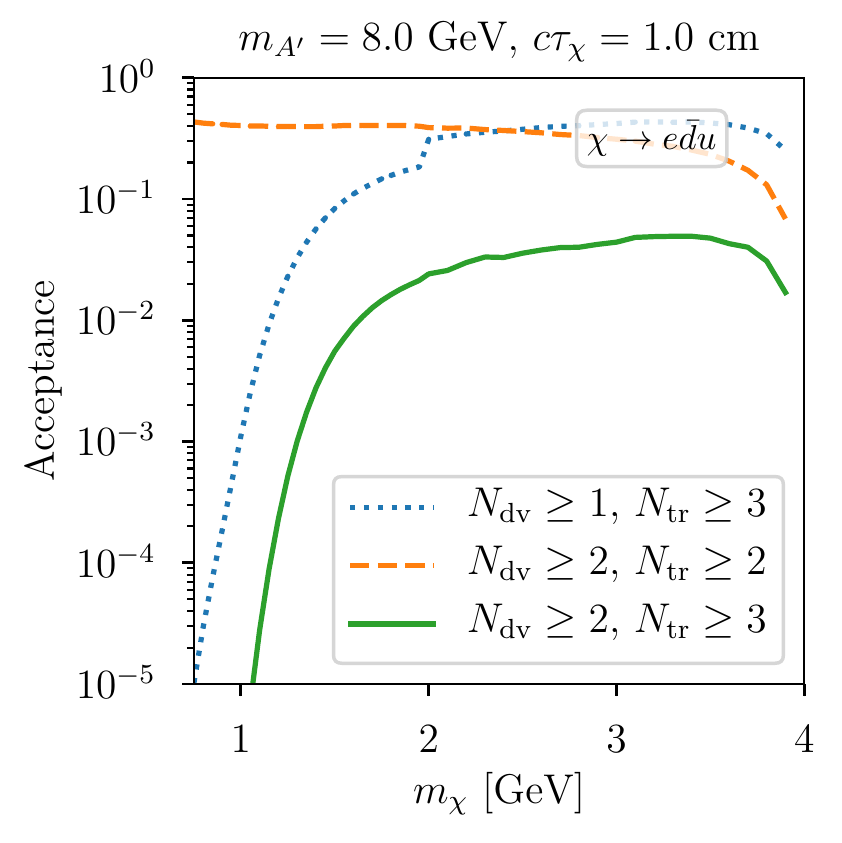}
            \includegraphics[width=0.47\textwidth]{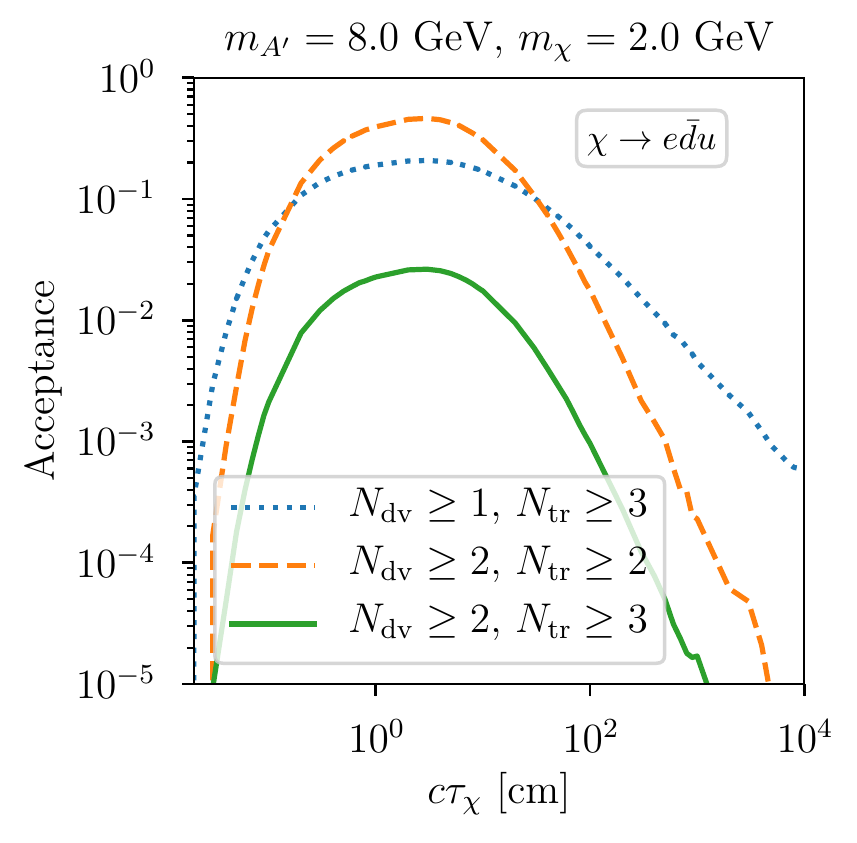}
      \caption{
        Slices of the event acceptance probability for fermion LLPs $\chi$ decaying to $e +\text{ hadrons}$ for dark photon production with $m_{A'}=8\;\GeV$. The 
        left (right) panel shows the acceptance as a function of the LLP mass for $c\tau_\chi = 1\;\cm$ 
        (as a function of $c\tau_\chi$ for $m_\chi = 2\;\GeV$.) We use the exclusive
        hadronic decay model of Appendix~\ref{sec:vector_ud_operator} for $m_\chi\le2\;\GeV$ and  use \pythia\ for showering and hadronization at larger $\chi$ masses; this explains the kink in the left plot at $2\; \GeV$.
      \label{fig:acceptance_slices_edu}}
    \end{figure}

\begin{figure}
      \centering
      \includegraphics[width=0.47\textwidth]{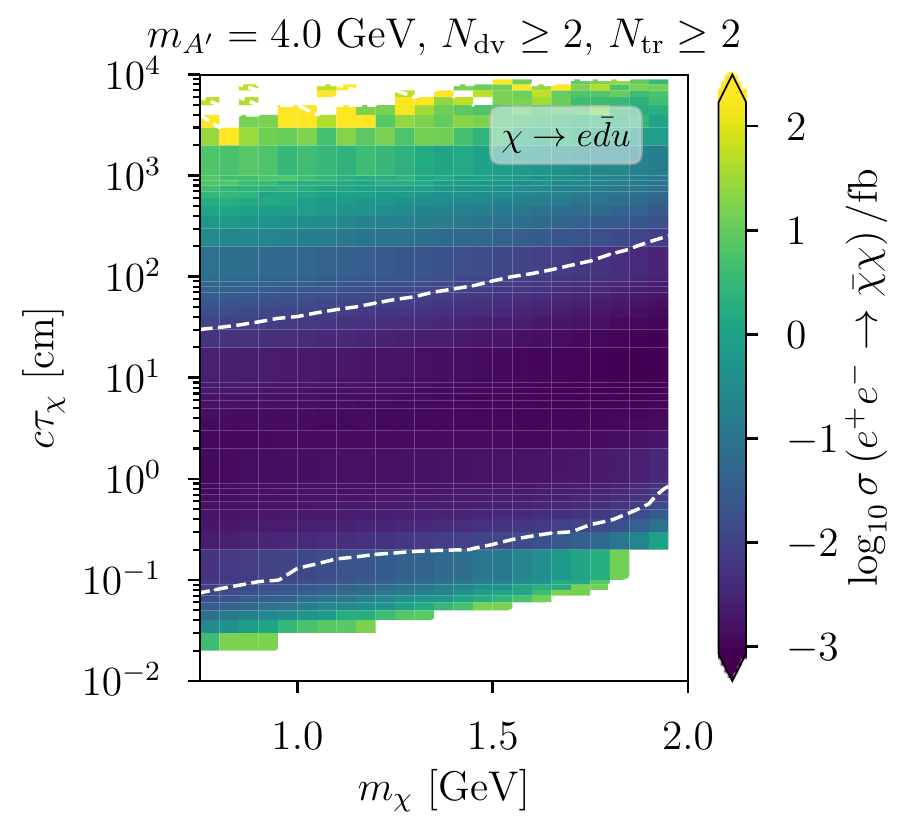}
      \includegraphics[width=0.47\textwidth]{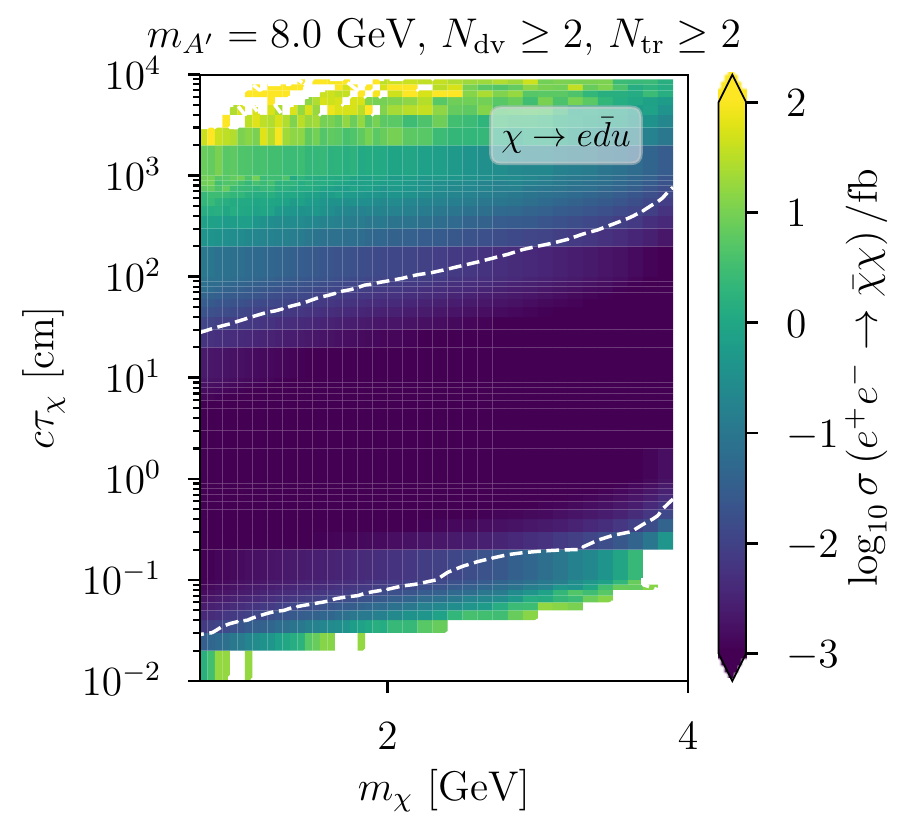} \\
      \includegraphics[width=0.47\textwidth]{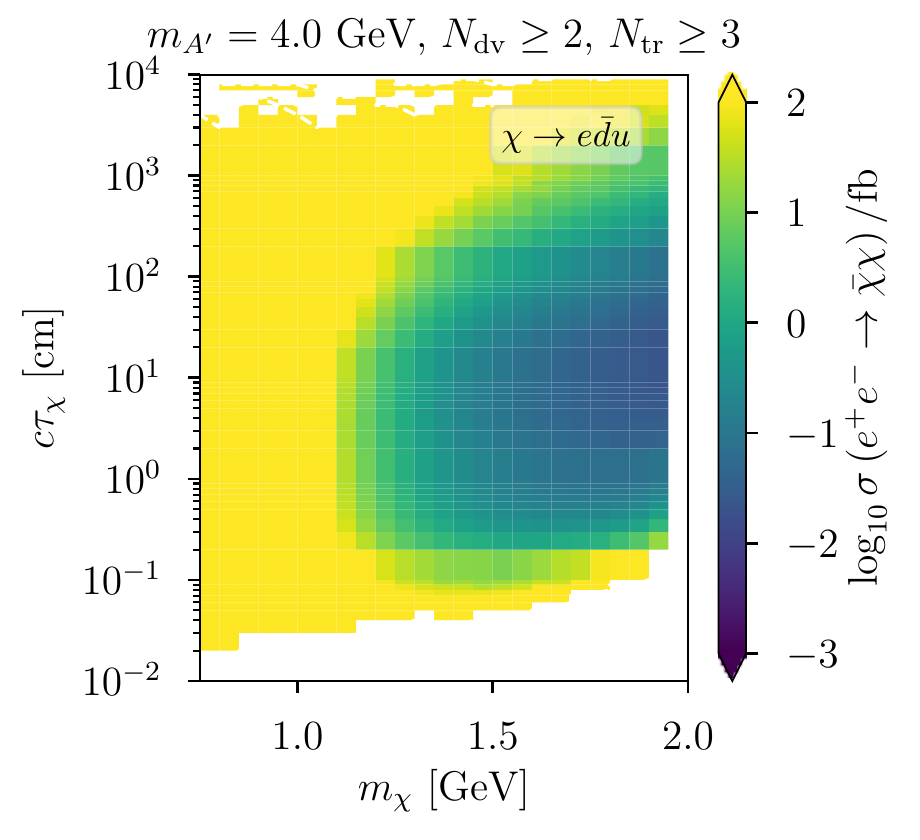}
      \includegraphics[width=0.47\textwidth]{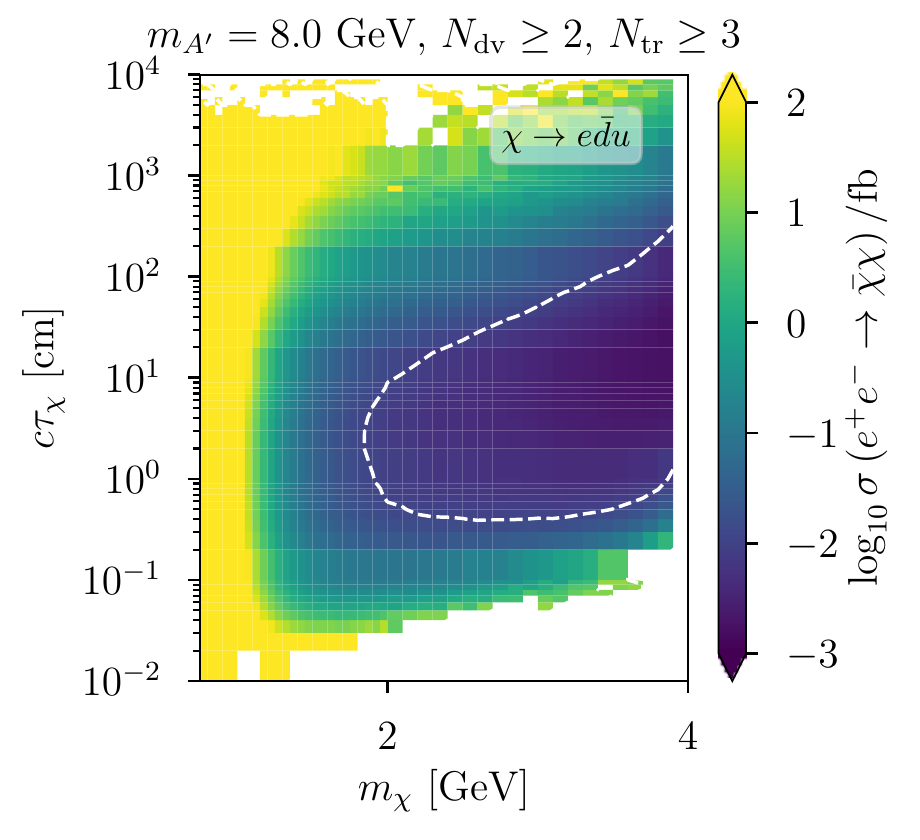} 
      \caption{
      Cross-section sensitivity in $\fb$ of Belle II based on 10 signal events in a $50\;\ab^{-1}$ dataset for 
      LLPs $\chi$ decaying to $e + \text{ hadrons}$ and the dark photon production mode. The upper (lower) row shows the result of 
        demanding at least 2 (3) charged tracks from two displaced vertices. The 
        left (right) column fixes $m_{A'} = 4$ ($8$) GeV. The white dashed contours 
        correspond to a production cross section of $10^{-2}\;\fb$, which is 
        roughly equivalent to $\varepsilon \sim 10^{-5}$ for $m_{A'} \leq 8\;\GeV$.
      \label{fig:xsec_sensitivity_edu}}
    \end{figure}

\begin{figure}
      \centering
      \includegraphics[width=0.47\textwidth]{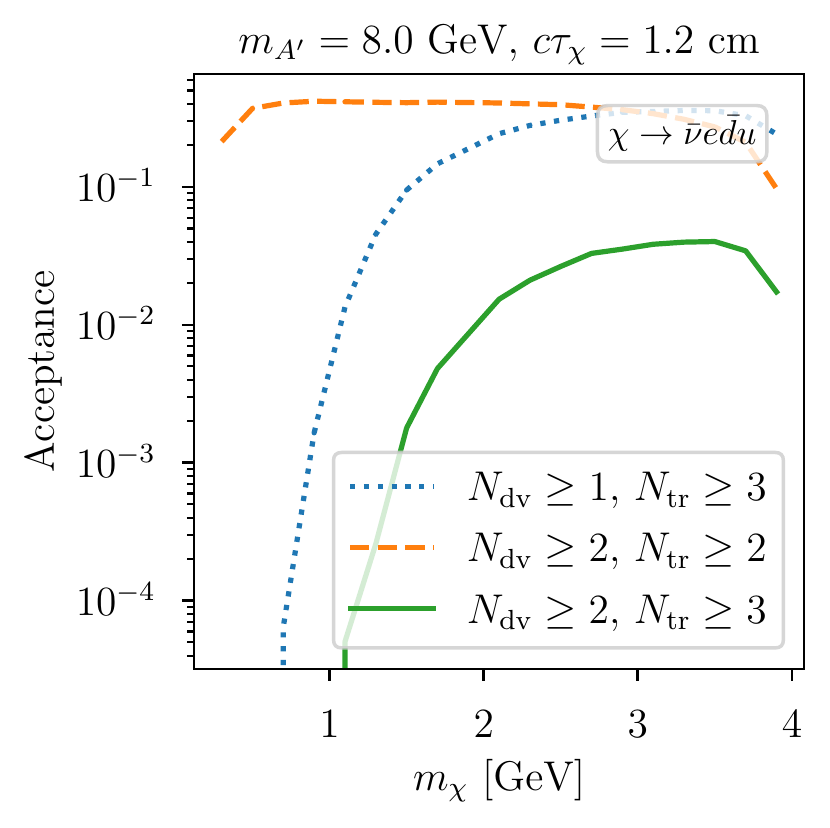}
      \includegraphics[width=0.47\textwidth]{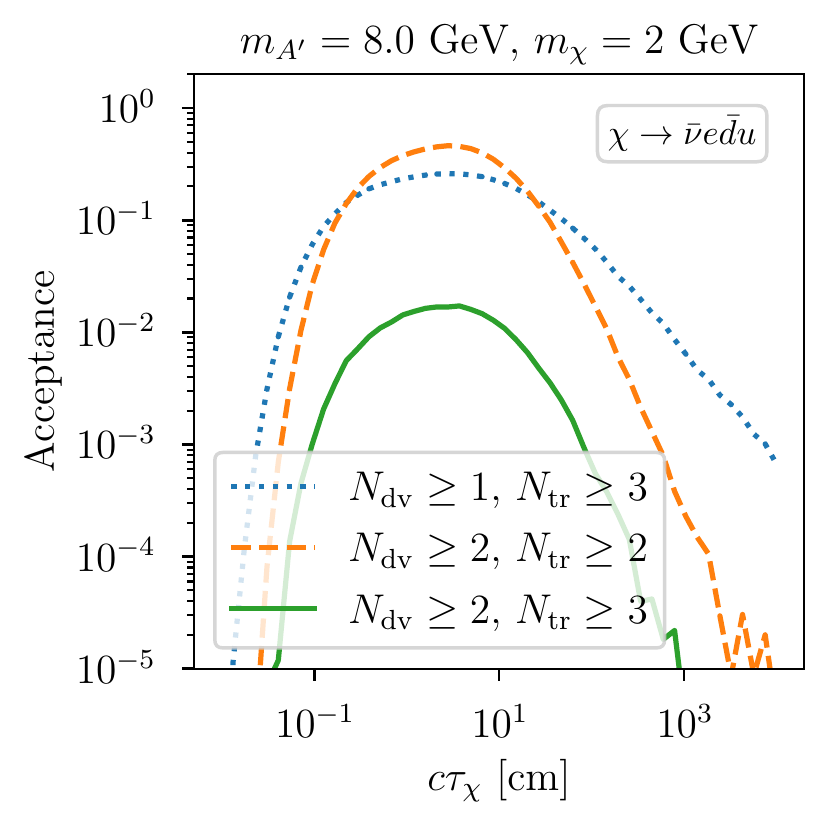}
      \caption{
        Slices of the event acceptance probability for LLPs $\chi$ decaying to $\nu e +\text{ hadrons}$ for dark photon production with $m_{A'}=8\;\GeV$. The 
        left (right) panel shows the acceptance as a function of the LLP mass for $c\tau_\chi = 1.2\;\cm$ 
        (as a function of $c\tau_\chi$ for $m_\chi = 2\;\GeV$.)
      \label{fig:acceptance_slices_nuedu}}
    \end{figure}

\begin{figure}
      \centering
      \includegraphics[width=0.47\textwidth]{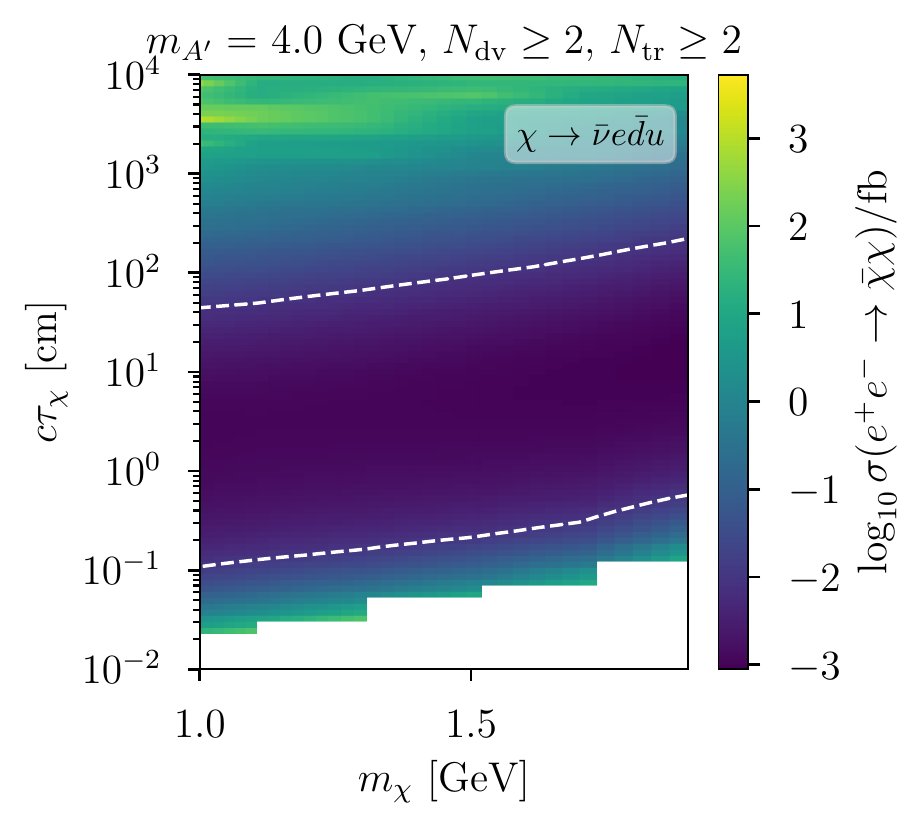}
      \includegraphics[width=0.47\textwidth]{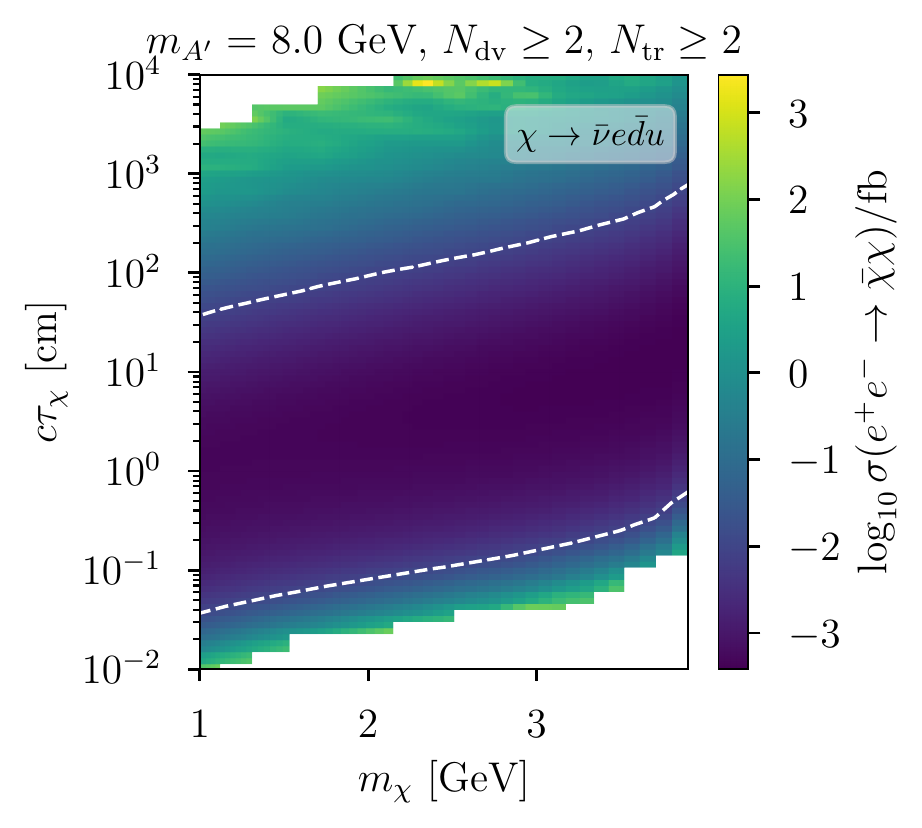} \\
      \includegraphics[width=0.47\textwidth]{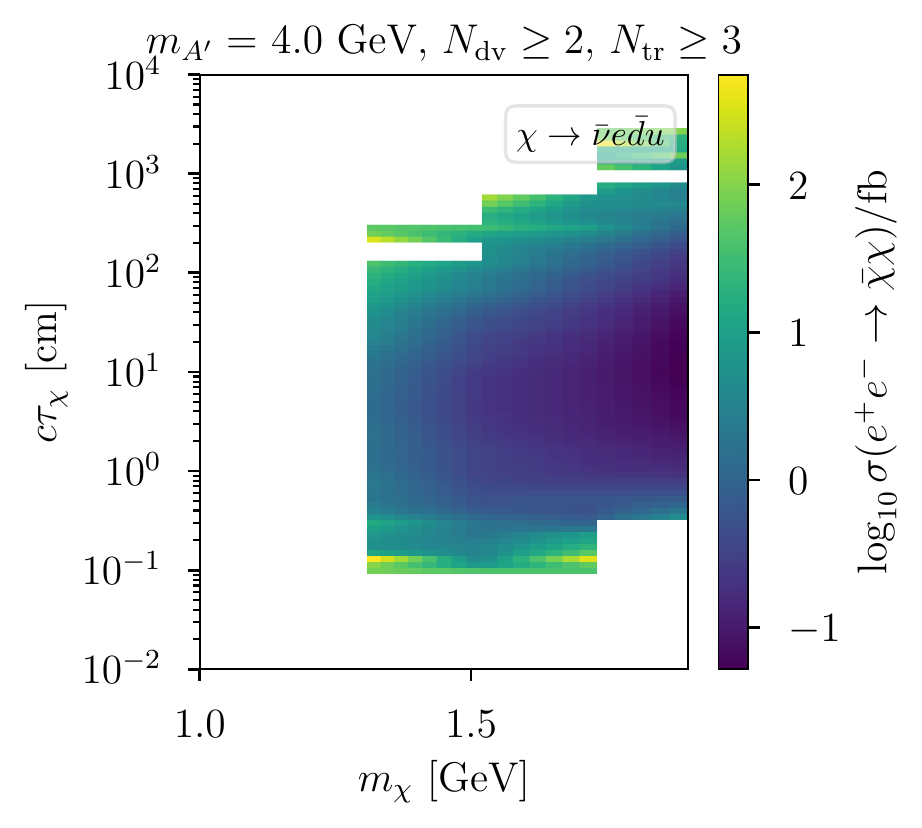}
      \includegraphics[width=0.47\textwidth]{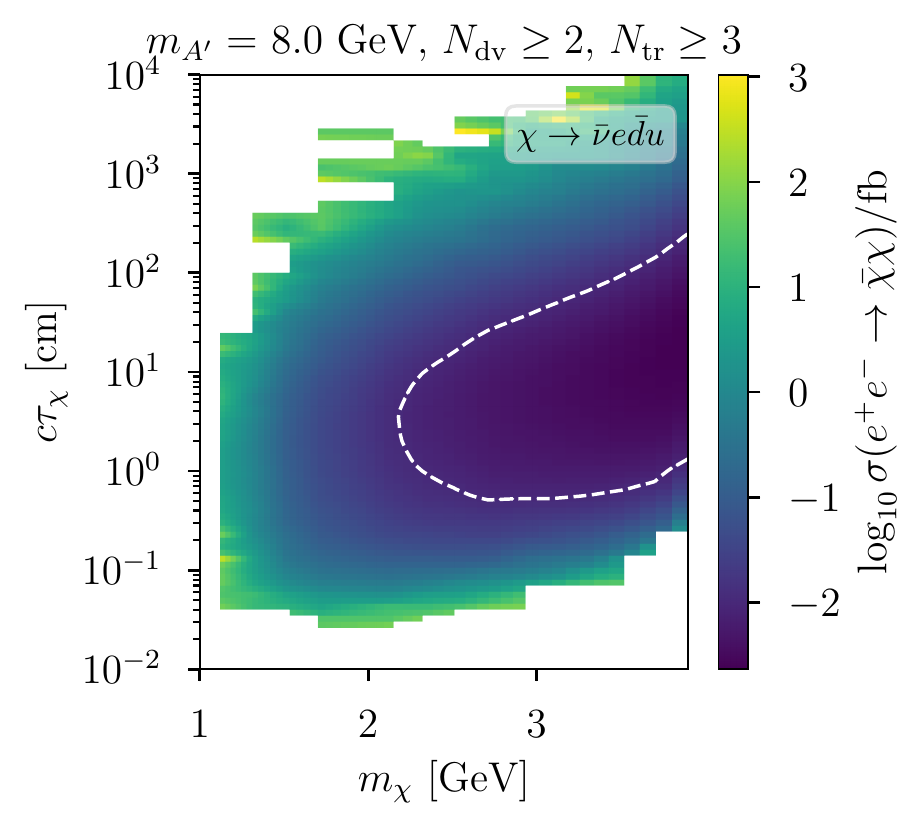} 
      \caption{
      Cross-section sensitivity in $\fb$ of Belle II based on 10 signal events in a $50\;\ab^{-1}$ dataset for 
      LLPs $\chi$ decaying to $\nu e + \text{ hadrons}$ and the dark photon production mode. The upper (lower) row shows the result of 
        demanding at least 2 (3) charged tracks from two displaced vertices. The 
        left (right) column fixes $m_{A'} = 4$ ($8$) GeV. The white dashed contours 
        correspond to a production cross section of $10^{-2}\;\fb$, which is 
        roughly equivalent to $\varepsilon \sim 10^{-5}$ for $m_{A'} \leq 8\;\GeV$.
      \label{fig:xsec_sensitivity_nuedu}}
    \end{figure}

\begin{figure}
      \centering
      \includegraphics[width=0.47\textwidth]{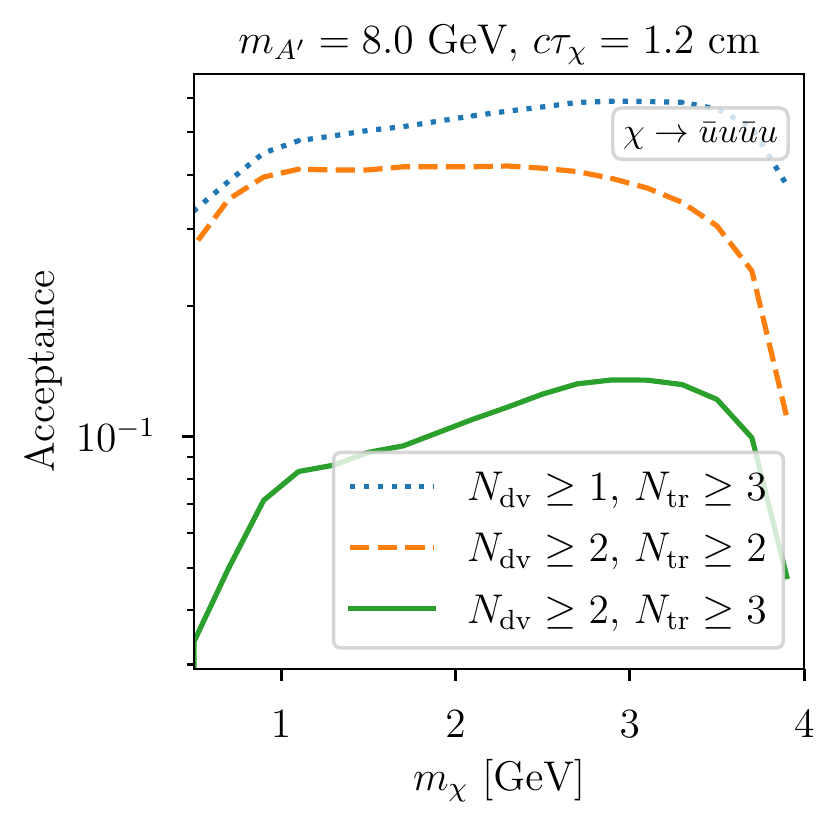}
      \includegraphics[width=0.47\textwidth]{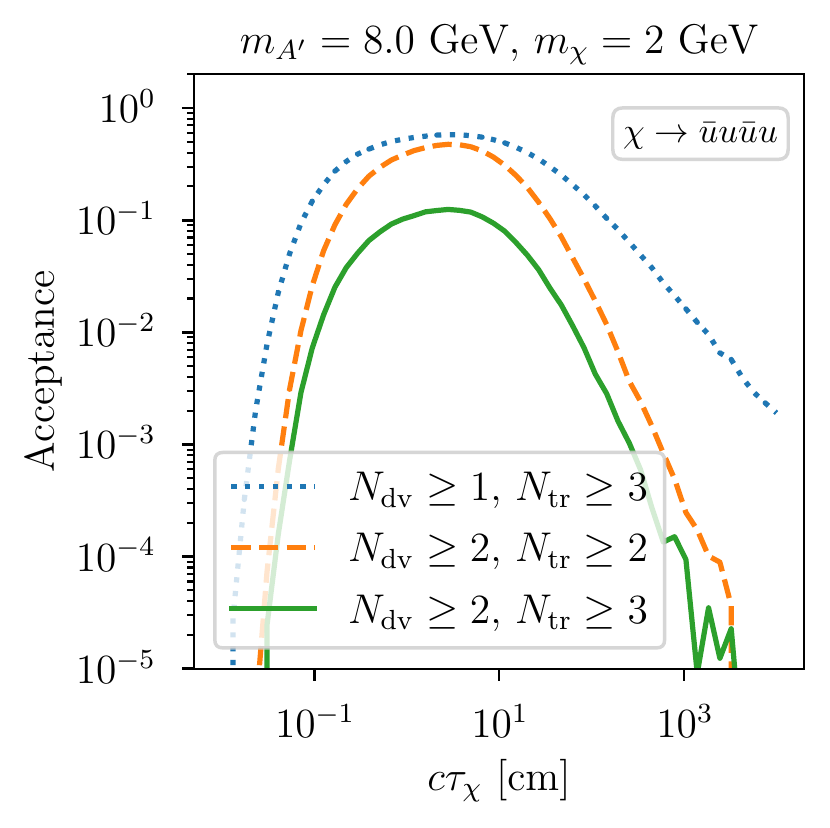}
      \caption{Slices of the event acceptance probability for LLPs $\chi$ decaying to hadrons for dark photon production with $m_{A'}=8\;\GeV$. The 
        left (right) panel shows the acceptance as a function of the LLP mass for $c\tau_\chi \approx 1\;\cm$ 
        (as a function of $c\tau_\chi$ for $m_\chi = 2\;\GeV$.) \label{fig:acceptance_slices_uuuu}}
    \end{figure}

\begin{figure}
      \centering
      \includegraphics[width=0.47\textwidth]{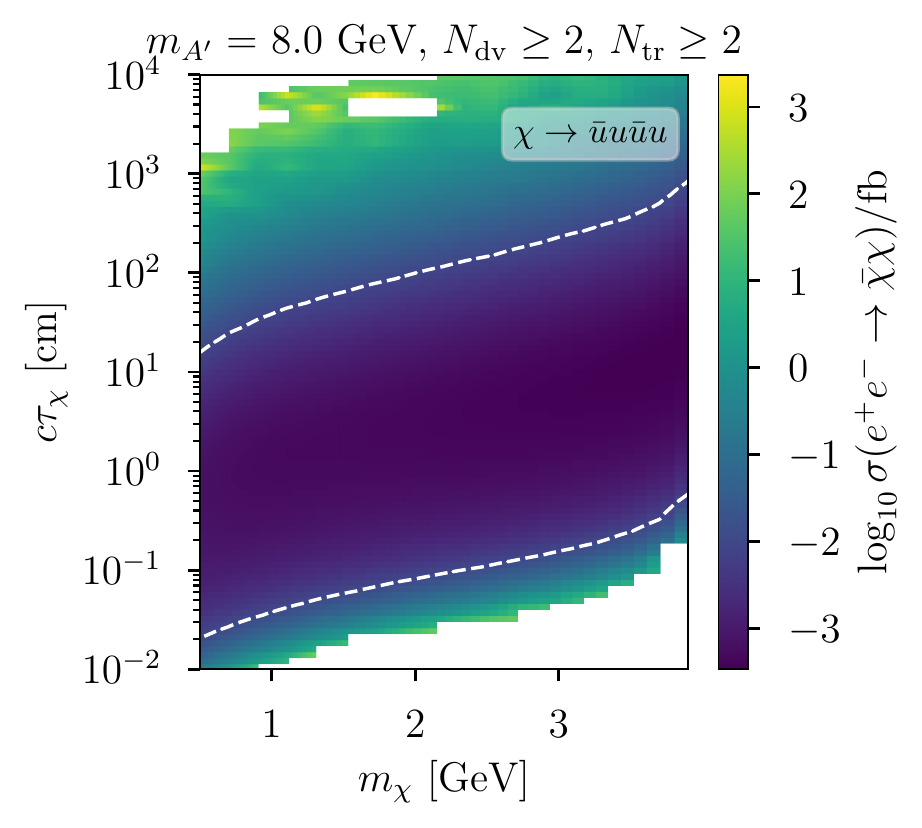} 
      \includegraphics[width=0.47\textwidth]{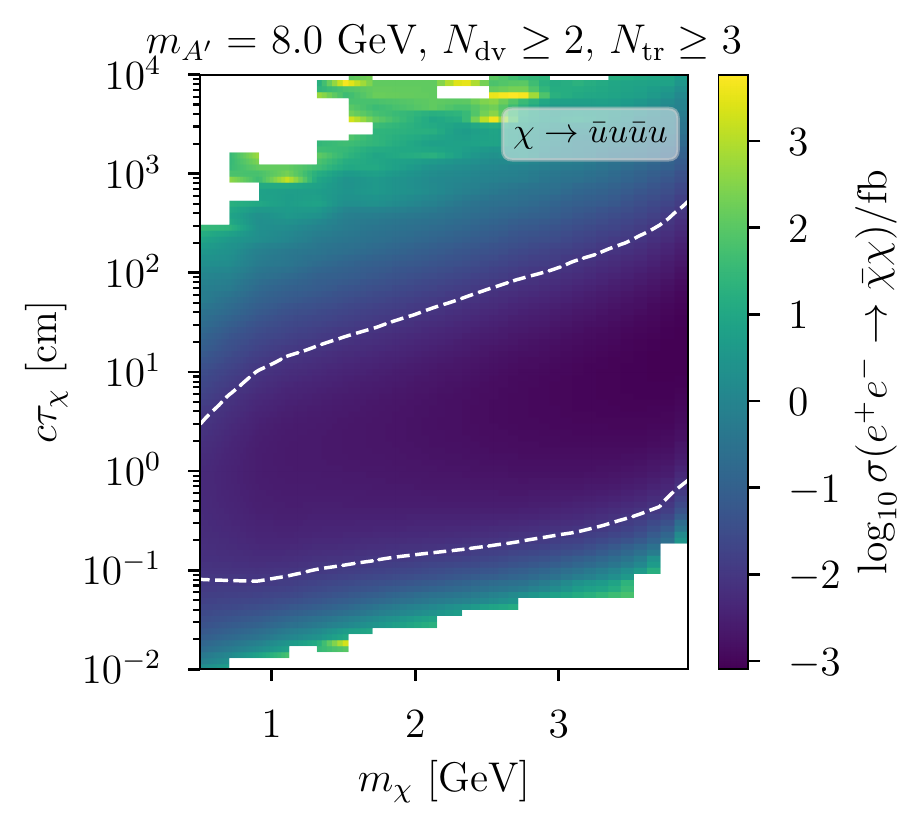} 
      \caption{
      Cross-section sensitivity in $\fb$ of Belle II based on 10 signal events in a $50\;\ab^{-1}$ dataset for 
        LLPs $\chi$ decaying to $\bar uu\bar u u$ and the dark photon production mode. The left (right) panel shows the result of 
        demanding at least 2 (3) charged tracks from two displaced vertices. We fix $m_{A'}=8$ GeV. 
        The white dashed contours 
        correspond to a production cross section of $10^{-2}\;\fb$, which is 
        roughly equivalent to $\varepsilon \sim 10^{-5}$ for $m_{A'} \leq 8\;\GeV$.
      \label{fig:xsec_sensitivity_uuuu}}
    \end{figure}

We first discuss the sensitivity of  multi-track displaced vertex searches to 
$e^+ e^- \to \gamma A^\prime$ with $A^\prime \to \chi \bar\chi$. 
The signal acceptance and cross-section sensitivity are shown for the following LLP decays, expressed in terms
of parton-level operators:
\begin{itemize}

\item Fully leptonic decay of scalar LLP $\chi \to \mu^+\mu^-e^+e^-$ in Figs.~\ref{fig:acceptance_slices_eemumu} and~\ref{fig:xsec_sensitivity_eemumu}; 

\item Semileptonic, fully visible decay of fermion LLP $\chi\to e \bar{d}u $ in Figs.~\ref{fig:acceptance_slices_edu} and \ref{fig:xsec_sensitivity_edu};

\item Semileptonic, partially invisible decay of scalar LLP $\chi \to \bar{\nu} e \bar{d} u$ in Figs.~\ref{fig:acceptance_slices_nuedu} and~\ref{fig:xsec_sensitivity_nuedu}; 

\item Fully hadronic decay of scalar LLP $\chi \to \bar{u}u\bar{u}u$ in Figs.~\ref{fig:acceptance_slices_uuuu} and~\ref{fig:xsec_sensitivity_uuuu}.

\end{itemize}

\paragraph{Fully leptonic LLP decays.} The most striking signals arise in fully leptonic decays. We consider for concreteness the decay of a scalar LLP, $\chi \to \mu^+\mu^-e^+e^-$; other flavour combinations are equally motivated. 
In this case, each event contains two vertices with four tracks and the acceptance is mainly determined 
by the LLP decay length. In the left panel of Fig.~\ref{fig:acceptance_slices_eemumu} we 
show the acceptance as a function of LLP mass for $c\tau_\chi = 1.2\;\cm$. The probability of 
observing two vertices is roughly the square of a single vertex, which is reflected in the relationship 
between the $\Ndv =1$ and $\Ndv =2$ acceptances. Finally, note that requiring additional tracks has a negligible 
impact on the acceptance, since each decay always produces four  tracks. A similar 
pattern is evident in the right panel of Fig.~\ref{fig:acceptance_slices_eemumu}, which shows the 
acceptance as a function of $c\tau_\chi$ for $m_\chi = 2\;\GeV$.

In Fig.~\ref{fig:xsec_sensitivity_eemumu} we estimate the sensitivity of 
the multi-track search for $m_{A'} = 4\;\GeV$ and $8\;\GeV$ in the left and right 
columns, respectively; the upper (lower) row shows analyses with $\Ndv \geq 2$, $\Ntr \geq 2$ ($\Ndv \geq 2$, $\Ntr \geq 3$).
Each panel shows the $10$ signal event cross-section reach in the $m_\chi - c\tau_\chi$ plane, 
with darker colours corresponding to smaller cross sections. The dashed white lines 
correspond to $\sigma = 10^{-2}\;\fb$, which translates to a kinetic mixing 
reach of $\varepsilon \sim 10^{-5}$; this can be compared with $\varepsilon \gtrsim 10^{-2}$ 
needed to explain the $(g-2)_\mu$ anomaly with $m_{A'}\gtrsim \;\GeV$. 
We see that our proposed searches offer excellent sensitivity independently of $\Ntr$ and $\Ndv$ selections. 
Moreover, the reach extends even to very low LLP masses $\lesssim 1\;\GeV$.

\paragraph{Semileptonic LLP decays.} The first semileptonic decay mode we consider is of a fermion LLP  $\chi\to e^-\bar{d} u$ via the $ \left(\bar u_{R}\gamma_\mu d_{R} \right)\left( \bar\ell_{R} \gamma^\mu \chi\right)$ operator. Like the fully leptonic decay above, this decay mode is fully reconstructible, although unlike the fully leptonic case the dominant decay modes for small LLP mass are $\chi\to e^-\rho^+$ and $\chi\to e^-\pi^+$, which feature only two charged tracks. However, at higher LLP mass the hadronization of $\bar d u$ can give rise to larger track multiplicities, especially due to intermediate vector resonances such as the $a_1^+$. We employ a $\tau$-inspired exclusive meson decay model outlined in Appendix~\ref{sec:vector_ud_operator} for $m_\chi\le2$ GeV, and \pythia\ for showering and hadronization for larger masses.\footnote{This transition point is chosen in the region where both hadronization approaches give comparable results (within a factor of two).} We show the acceptances for this model in Fig.~\ref{fig:acceptance_slices_edu}. The acceptances for the $\Ndv\ge2$, $\Ntr\ge3$ search are reduced by over an order of magnitude relative to the leptonic case due to the reliance on hadronization to produce extra tracks. The acceptance for this search is greatly reduced at low $m_\chi$ because the dominant $\chi$ decays have two tracks in this mass range. As before, the acceptance of $\Ndv \geq 2$ events 
is roughly the square of the $\Ndv\geq 1$ case. 

    Although the acceptances are smaller than for fully leptonic decays, they are still not negligible. The cross-section sensitivities for several dark photon mass benchmarks are shown in Fig.~\ref{fig:xsec_sensitivity_edu}. While the $\Ndv\geq 2$, $\Ntr \geq 2$ sensitivity shown in the upper panels is qualitatively 
similar to the fully leptonic case, the $\Ntr \geq 3$ search is limited at low LLP masses 
as explained above. The signal acceptances for lower $m_{A'}$ masses and $\Ntr\geq3$ (lower left panel) are more negatively affected since 
$m_\chi$ is restricted to smaller values such that the $A'\rightarrow \bar\chi\chi$ is still kinematically allowed. Despite this, we see that the multi-track search still offers sensitivity to 
tiny cross sections across a broad range of parameter space.
    
Next, we consider an even more challenging LLP semileptonic final state:~$\chi \to \bar\nu e \bar d u$.
For this LLP decay mode, an inclusive search strategy is required because the decay is not fully reconstructible.
The signature
includes  missing energy in addition to the lower track multiplicity from $\bar d u$ and,
once again, these challenges are clearly reflected in the acceptances in Fig.~\ref{fig:acceptance_slices_nuedu}. 
The resulting search reach is shown in Fig.~\ref{fig:xsec_sensitivity_nuedu}. However, the results are qualitatively
similar between the two semileptonic $\chi$ decay modes.  Note that there are larger theoretical uncertainties on the
hadronization of this scalar operator compared to the vector $\bar{u}_{R}\gamma^\mu d_{R}$ operator above. In 
Figs.~\ref{fig:acceptance_slices_nuedu} and \ref{fig:xsec_sensitivity_nuedu} we use \pythia for showering and hadronization, and 
we provide an alternative 
model based on chiral perturbation theory (as well as comparisons with \pythia) in Appendix \ref{sec:scalar_du_operator}. 

\paragraph{Fully hadronic LLP decays.} For completeness, we also consider fully hadronic LLP decays.
As a benchmark, we study the final states with $\chi \to \bar u u \bar u u$. 
Hadronization of this final state produces a pair of neutral pions most of the time, so the requirement 
of multiple tracks per vertex can penalize the acceptance. Branching fractions into charged states 
is also very sensitive to the hadronization model, so our predictions for these decays are subject to a 
large systematic uncertainty. We emphasize that this does not affect the experiment's ability to carry out the proposed searches.
In Fig.~\ref{fig:acceptance_slices_uuuu} we show the acceptance as function of LLP mass in the left panel and as a function 
of its lifetime in the right panel for event samples generated using \pythia to hadronize the final states.
As we show in Appendix~\ref{sec:hadronization}, this most likely provides an optimistic estimate of the acceptance, particularly 
at low LLP masses $\lesssim 2\;\GeV$. Fig.~\ref{fig:xsec_sensitivity_uuuu} shows the cross-section sensitivity for the same hadronization 
procedure. As in the previous examples, multi-track searches can be sensitive to cross sections far smaller than those 
motivated by solutions to $(g-2)_\mu$. Note that we do not show the $m_{A'} = 4\;\GeV$ case because 
of large hadronization uncertainties for $m_\chi \leq m_{A'}/2 = 2\;\GeV$.


    \subsection{Dark Higgs Production in \texorpdfstring{$B$}{B} Meson Decays}
We now turn to LLP production in the decays of a dark Higgs. The full process is $B\to K\phi,\phi\to\bar\chi\chi$, where we sum over various kaon resonances. The primary effect of the change in production mode is that it alters the boost of the LLPs, which shifts the values of $c\tau_\chi$ corresponding to the optimal decay length sensitivity of 1 mm--50 cm for a tracker-based displaced vertex analysis. However, the acceptances are otherwise uncharged, largely because the track $p_T$ selections are easily passed independent of the LLP boost for the $\chi$ masses we consider.

Because the sensitivities of our proposed displaced vertex searches are largely unchanged from the dark photon production mode, we only show results for one of the LLP decay modes, namely the fermion LLP $\chi\to e\bar du$ decay originating from the $ \left(\bar u_{R}\gamma_\mu d_{R} \right)\left( \bar\ell_{R} \gamma^\mu \chi\right)$ operator. We have selected this decay mode for illustrative purposes and because of the robustness of our modelling of the hadronic part of the decay over a wide mass range; however, we emphasize that all possible $\chi$ decay modes allowed by the EFT are comparably motivated.

      \begin{figure}
      \centering
      \includegraphics[width=0.47\textwidth]{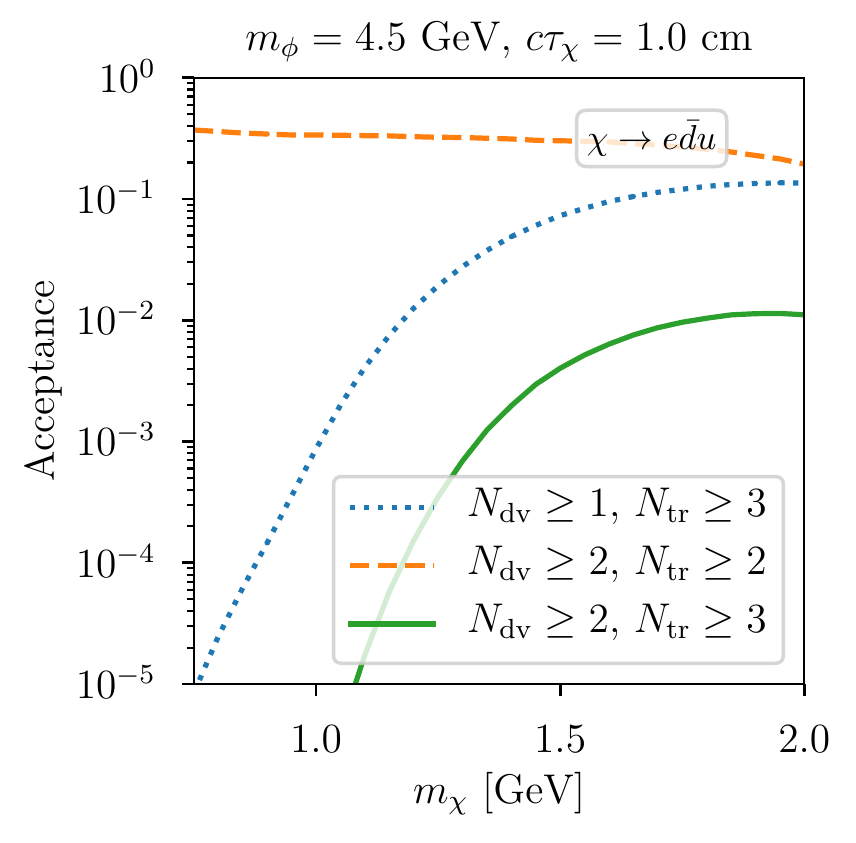}
      \includegraphics[width=0.47\textwidth]{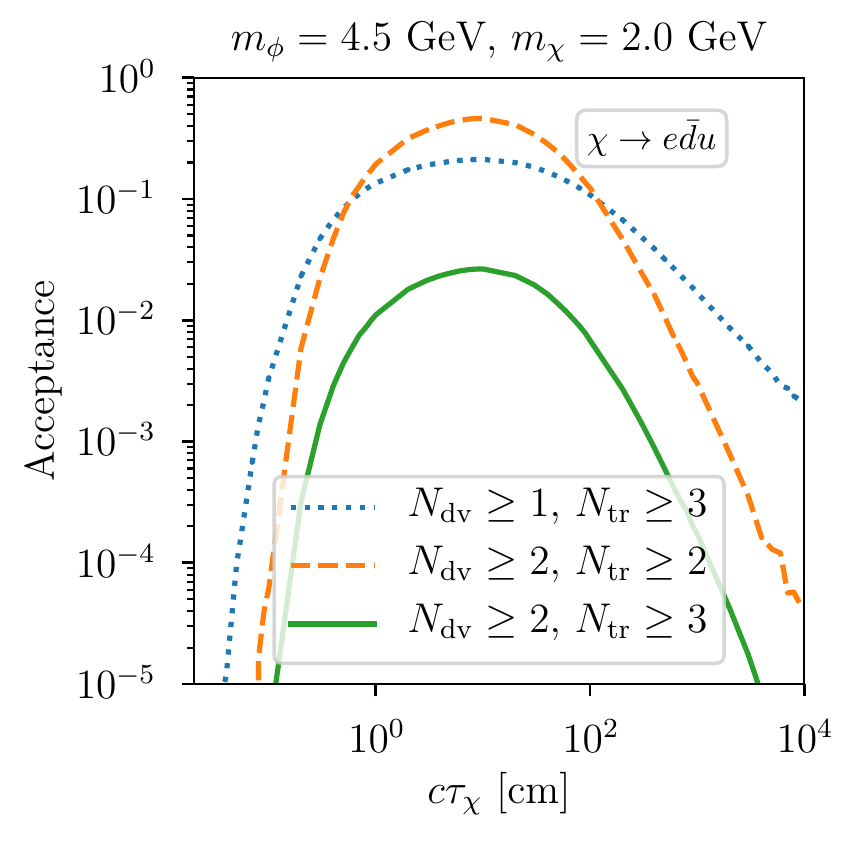}
      \caption{
        Slices of the event acceptance probability for LLPs $\chi$ decaying to $ e +\text{ hadrons}$ for the
         $B\to K\phi,\phi\to\bar\chi\chi$ production mode and $m_\phi=4.5\;\GeV$. The 
        left (right) panel shows the acceptance as a function of the LLP mass for $c\tau_\chi=1\;\cm$ 
        (as a function of $c\tau_\chi$ for $m_\chi = 2\;\GeV$.)
      \label{fig:acceptance_slices_Bdecay_edu}}
    \end{figure}

\begin{figure}
      \centering
      \includegraphics[width=0.47\textwidth]{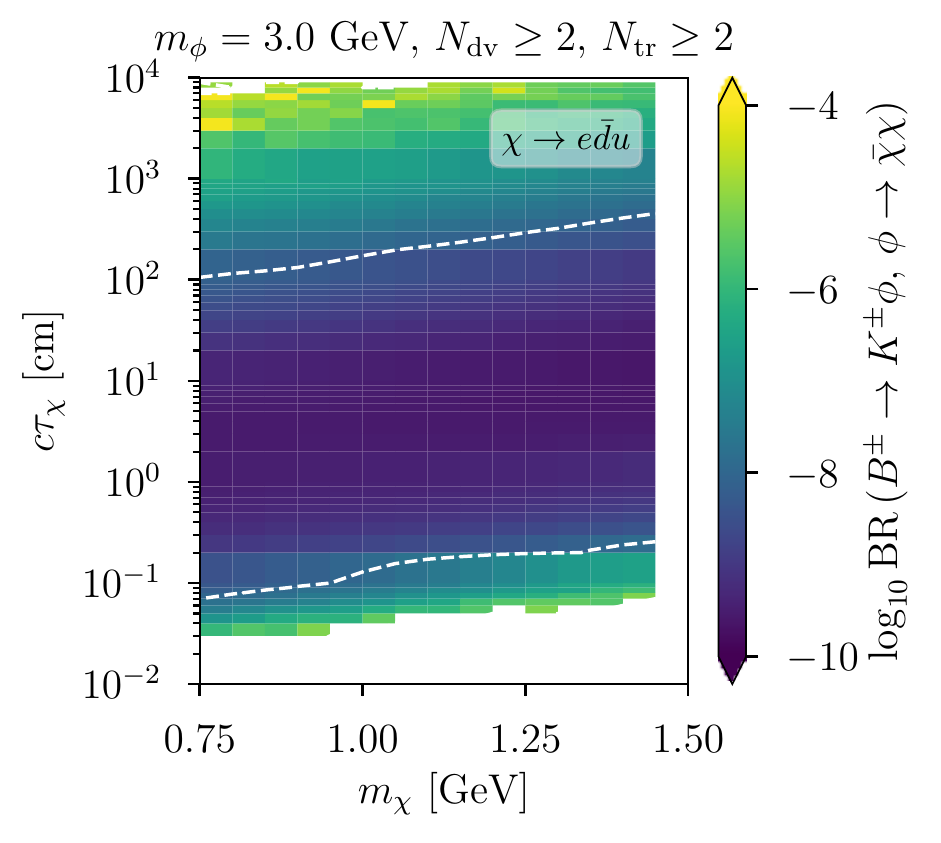}
      \includegraphics[width=0.47\textwidth]{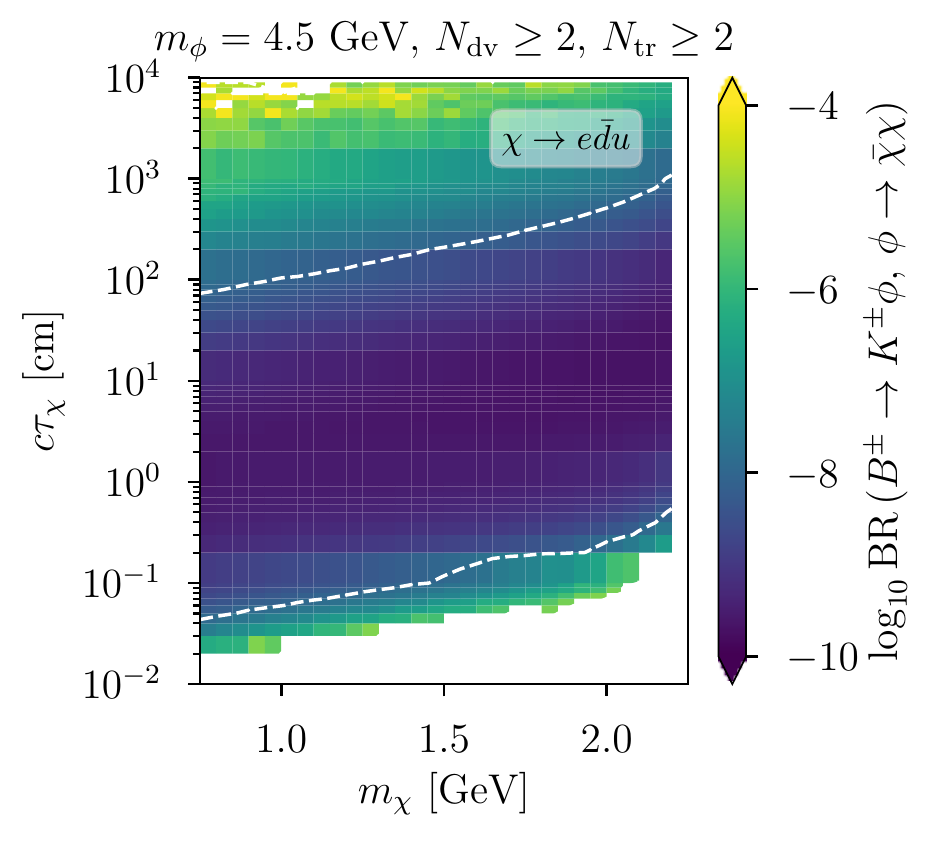} \\
      \includegraphics[width=0.47\textwidth]{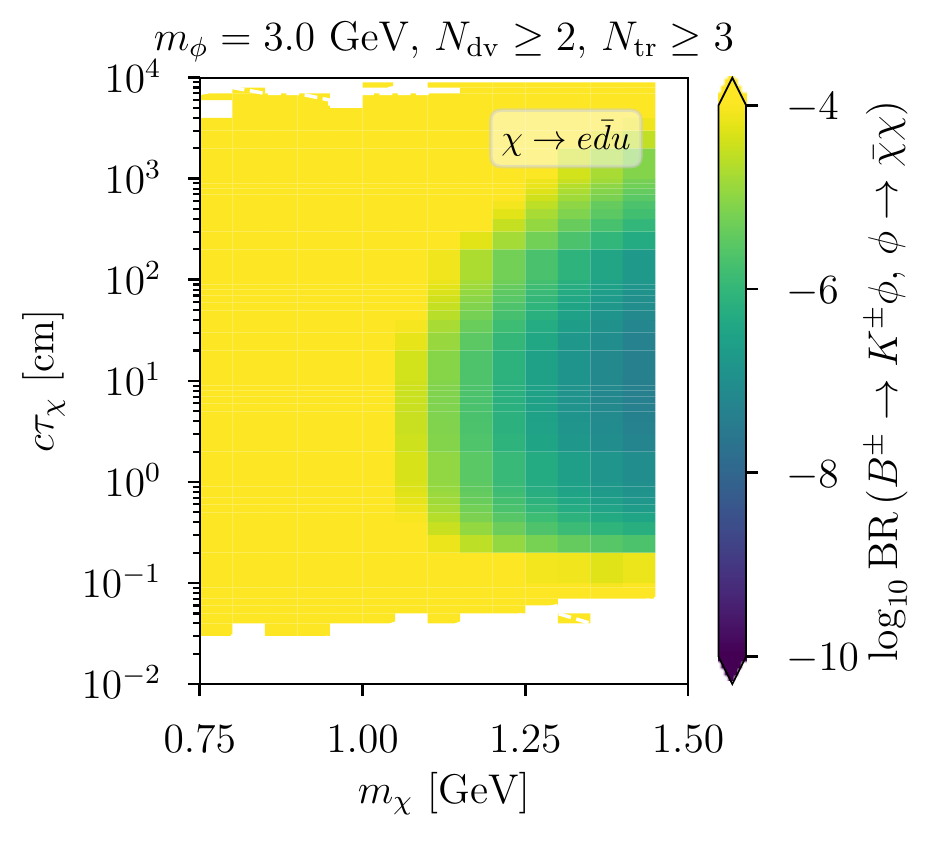}
      \includegraphics[width=0.47\textwidth]{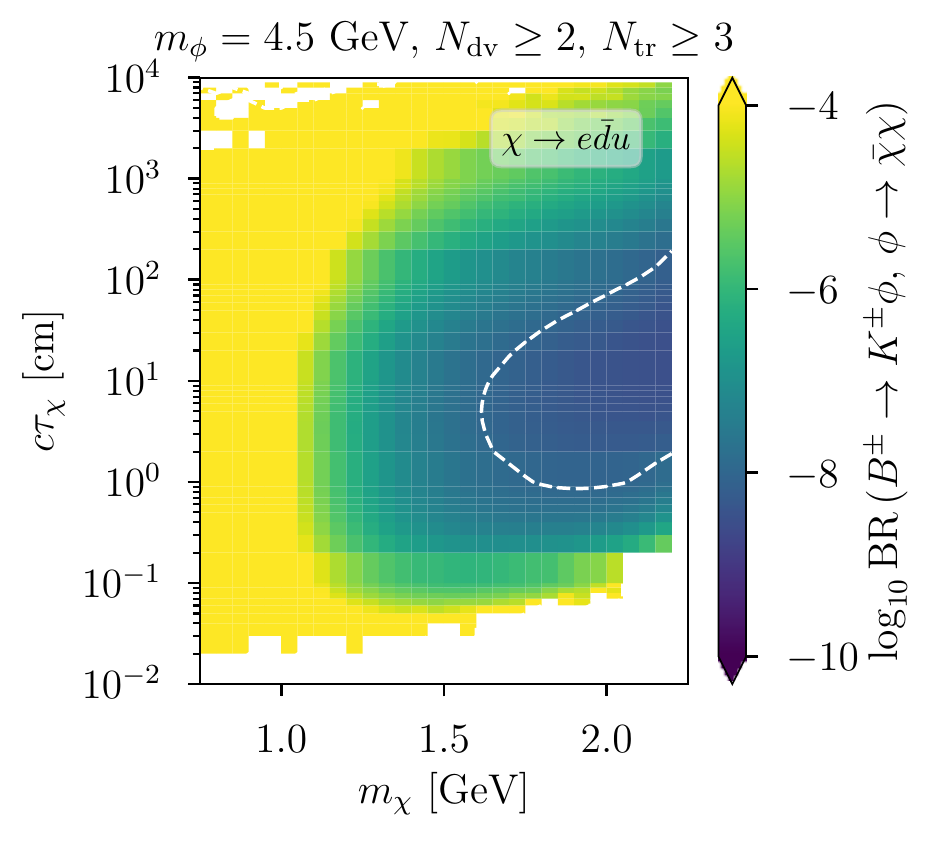} 
      \caption{
      $B$ meson branching fraction sensitivity of Belle II based on 10 signal events in a $50\;\ab^{-1}$ dataset for 
      LLPs $\chi$ decaying to $e + \text{ hadrons}$ and the $B$ decay production mode (summed over available kaon
      final states). The upper (lower) row shows the result of 
        demanding at least 2 (3) charged tracks from two displaced vertices. The 
        left (right) column fixes $m_{\phi} = 3$ ($4.5$) GeV. The white dashed contours 
        correspond to a production cross section of $10^{-2}\;\fb$, which is 
        roughly equivalent to dark Higgs mixing $\theta \sim 10^{-4}$ for $m_{\phi} \leq 4.5\;\GeV$.
      \label{fig:xsec_sensitivity_Bdecay_edu}}
    \end{figure}

In Fig.~\ref{fig:acceptance_slices_Bdecay_edu}, we show the acceptances for various requirements on $\Ndv$ and $\Ntr$ for the $B$ decay production mode. Most striking is the fact that the right pane of Fig.~\ref{fig:acceptance_slices_Bdecay_edu} is nearly identical to the right pane of Fig.~\ref{fig:acceptance_slices_edu}, which shows the acceptance dark photon production and the same decay mode. The value of $c\tau_\chi$ for which the acceptance peaks is shifted between the two production modes, reflecting the different characteristic boosts for the LLPs, but the curves are otherwise very similar. We show in Fig.~\ref{fig:xsec_sensitivity_Bdecay_edu} the sensitivity to the branching fraction of $B^\pm\to K^\pm\phi$ assuming a 100\% branching fraction of $\phi\rightarrow\bar\chi\chi$, finding once again similar sensitivity to the dark photon production mode in terms of cross section. We again show the 10-event sensitivity to a  cross section of $10^{-2}\;\fb$, which in the dark Higgs model corresponds to a mixing with the SM Higgs of $\theta\sim10^{-4}$ for $m_\phi=4.5\;\GeV$. This analysis demonstrates that sufficiently inclusive searches can have comparable acceptances regardless of LLP production mode.

\section{Backgrounds}
\label{sec:backgrounds}

Backgrounds for LLP searches are notoriously difficult to estimate \cite{Alimena:2019zri}, particularly in the absence of access to a full detector simulation. Data-driven methods are also needed to model extremely rare processes that may not be captured in simulations. In this section, we do not attempt to accurately model backgrounds for LLP searches at Belle II, but instead we  enumerate the main expected sources of background, and discuss how they can be characterized and mitigated in multi-track displaced vertex searches. As a general rule, the backgrounds to LLP searches are very rare, and so increasing either the multiplicity of vertices or the number of tracks per vertex can  suppress them.

The backgrounds fall into three main categories. First, there are backgrounds from fake vertices and tracks, whether arising from collisions of primary particles with gas or material in the detector, accidental track crossings, or spurious tracks that arise from mis-reconstruction of tracker hits. Second, there are backgrounds from heavy-flavour decays such as $b$ and $c$ hadrons, which mimic the signal by having relatively large masses and high track multiplicities. Finally, there are backgrounds from hadrons with strange quarks, such as kaons and $\Lambda$ baryons. While these decays typically give low track multiplicities and can often be fully reconstructed and rejected, they are produced in sufficient numbers that rare decays or mis-reconstruction of the vertices could populate the signal region. All masses, lifetimes, and branching fractions of SM particles are taken from the Particle Data Group world averages~\cite{Zyla:2020zbs}.

\subsection{Combinatoric and Detector Backgrounds}

Displaced vertices can originate from particle interactions with the detector material, such as photon conversions into $e^+e^-$ pairs, as well as random overlaps of high-impact-parameter tracks from the decays of kaons, pions, and strange baryons. Spurious vertices from material interactions can be removed by vetoing vertices that originate in material-dense regions of the detector. Additionally, photon conversions and material interactions should typically give rise to low multiplicities and small invariant masses of the tracks originating from the vertex, and so increasing the $\Ntr$ requirement and applying a modest cut on the vertex track invariant mass can remove these vertices.

More problematic are random crossings or mis-reconstructed tracks that are accidentally combined into a displaced vertex. In the \babar\ search for $\Ntr=2$ vertices \cite{Lees:2015rxq}, this constituted the dominant background. Based on the total \babar\ trigger rate and the number of background events, we calculate the probability of a fake single, two-track vertex to be approximately $10^{-6}$ for that analysis, which would correspond to $10^{5}$ such vertices with the full Belle II integrated luminosity. In reality, the fake-vertex background will be larger because of the larger beam backgrounds at Belle II. Nevertheless, random-crossing vertices become rarer for larger vertex multiplicities and track multiplicities per vertex, and so requiring $\Ndv\ge2$ and/or $\Ntr\ge3$ should render this background negligible. Fortunately, the fake vertex rate can be estimated from data:~for example, ATLAS and CMS pair high-impact-parameter tracks and vertices from different events to efficiently estimate the magnitude of this background in LHC searches \cite{Aad:2015rba,Sirunyan:2021kty}. A similar approach at Belle II would likely be effective.

\subsection{Heavy-Flavour Backgrounds}

Heavy-flavour backgrounds, such as $B$ mesons, $D$ mesons, and $\tau$ leptons, are pernicious because they give rise to genuine high-mass,  multi-track displaced vertices. Their typical decays cannot be fully reconstructed, which means they cannot be readily vetoed. Fortunately, the lab-frame decay lengths of these particles at Belle II are sub-mm:~$\beta\gamma c\tau$ is approximately  0.14 mm for $B$ mesons ($\beta\gamma\sim0.3$) and 0.26 mm for $\tau$ leptons ($\beta\gamma\sim3$). $D$ mesons have different kinematics depending on whether they originate from continuum $c\bar{c}$ production or $B$ meson decay, but we expect a typical $\beta\gamma\sim1$ (with a maximum value of 3), which corresponds to a lab-frame decay length of 0.3 mm (0.9 mm).

With the illustrative minimum vertex displacement we used in Sec.~\ref{sec:analysis} of 2 mm, we expect to reduce $B$ backgrounds by $2\times10^{-7}$, $\tau$ backgrounds by $4.5\times10^{-4}$, and charm backgrounds by approximately $10^{-3}$ (although this depends on how many charm mesons populate the highest-momentum tail). Increasing the minimum vertex displacement to  4 mm could reduce even charm backgrounds to the $10^{-6}$ level, at which point requiring $\Ndv\ge2$ would effectively suppress heavy-flavour backgrounds. Tightening the minimum vertex displacement would eliminate some sensitivity to signals with the shortest values of $c\tau$, but would otherwise have a negligible impact on signal acceptance. Because heavy-flavour backgrounds follow an exponential distribution in decay length and have kinematics that can be well-measured in other processes, they can be readily estimated by looking at sidebands that are just below the signal region in displacement.

While raising the minimum required vertex displacement can effectively suppress heavy-flavour backgrounds, it may be possible to mitigate them using other means depending on the signal being searched for. As a result, sensitivity to smaller signal lifetimes could be possible at the expense of making the searches less inclusive. Examples of possible strategies could include requiring multiple leptons per displaced vertex, or (in the case of charm backgrounds) vetoing vertices with kaons. The leading multi-charged-pion decay modes of $D$ mesons (such as $D^\pm\rightarrow \pi^+\pi^-\pi^-$) occur at the $10^{-2}$ level, compared to $\ge20\%$ for 3-track decays including kaons. In $\tau$ events, there should be a very low multiplicity of particles that are unassociated with the displaced vertex, since most taus decay to a single track plus missing momentum. $B$ backgrounds can be suppressed using event-shape observables like thrust \cite{Farhi:1977sg} and Fox-Wolfram moments \cite{Fox:1978vu}, which can identify the characteristic kinematics and event topologies associated with $B$ mesons; however, if this strategy were pursued, it would be important to test whether this would also limit sensitivity to LLPs produced in $B$ decays, such as $B\rightarrow K\phi,\,\phi\rightarrow \chi\bar\chi$.

\subsection{Strange-Quark Backgrounds}

Finally, we turn to strange-quark backgrounds such as $K^+$, $K_S^0$, $K_L^0$, and $\Lambda$. These typically feature low-track-multiplicity decays, but have sufficiently long lifetimes that they frequently give displaced decays inside of the signal region.

\begin{figure}
  \centering
  \includegraphics{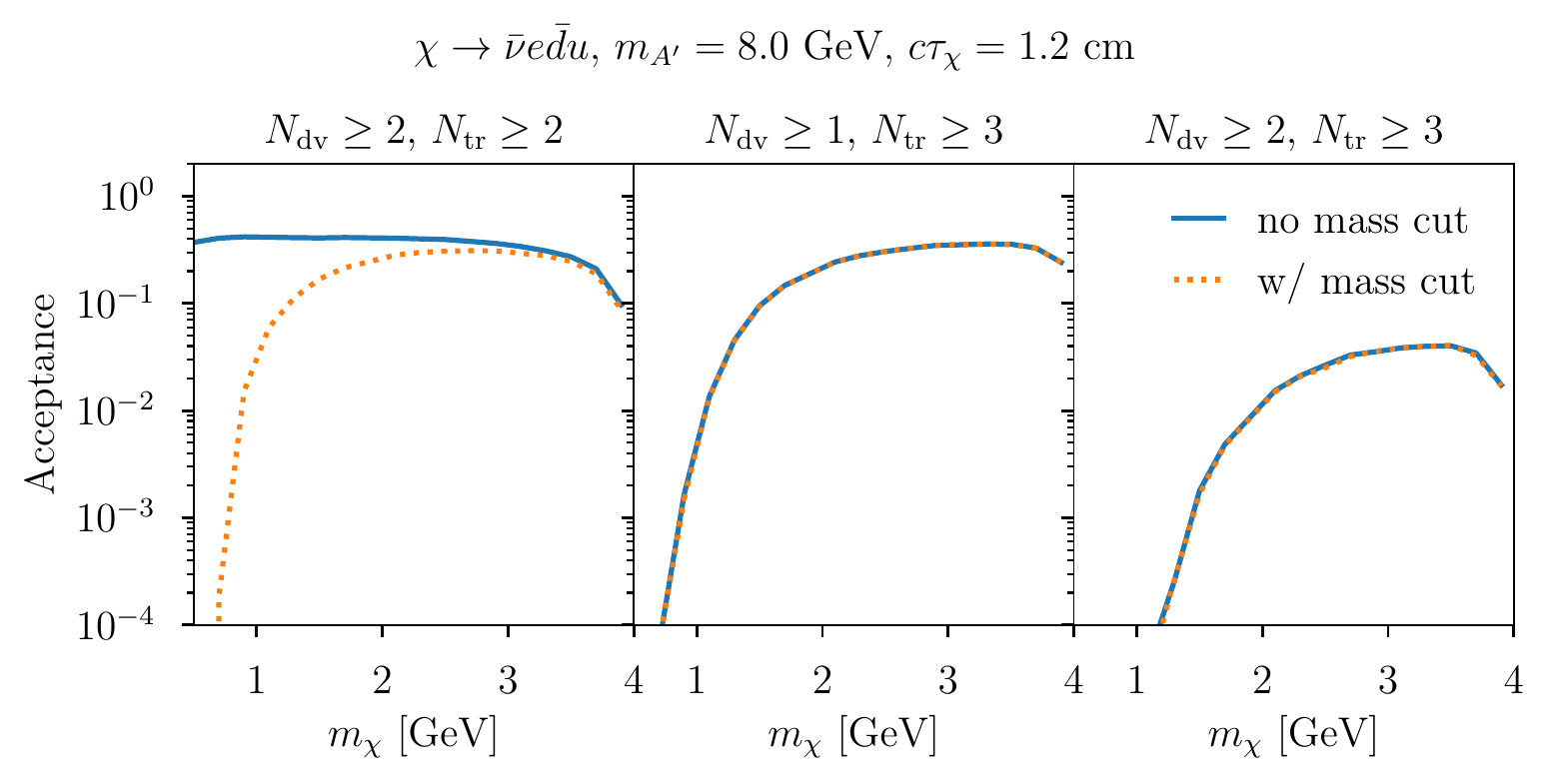}
  \caption{Comparison of signal acceptance with and without  a vertex track-mass cut $> 0.5\;\GeV$ to reduce 
  kaon backgrounds.\label{fig:trackmass_acceptance}}
\end{figure}

\paragraph{Charged kaons.} With $c\tau\approx 0.4$ m, most $K^\pm$ are stable through the tracking system. Kaons are typically produced with an $\mathcal{O}(1)$ boost, which means that we expect 1\%--10\% of charged kaons to decay in the signal region. Given that we expect about one kaon per hadronic event \cite{Lees:2013rqd}, this translates to $\sim10^{11}$ kaons with the full Belle II integrated luminosity. The most common, dangerous decay mode is $K^+\rightarrow \pi^+\pi^+\pi^-$ with branching fraction of 5.5\%. This is, however, a fully reconstructible decay and can be vetoed:~if the veto has an inefficiency of $10^{-3}$ for $K^\pm$, this should be sufficient to render the background negligible for $\Ndv\ge2$. Furthermore, tightening the track requirement to $\Ntr\ge4$ would eliminate these backgrounds while having little effect on neutral signal DVs. Other problematic decay modes include $K^+\rightarrow \pi^+\pi^- e^+\nu_e$ with a branching fraction of $4\times10^{-5}$, $K^+\rightarrow \pi^+\pi^+\pi^-\gamma$ with a branching fraction of $7\times10^{-6}$, and $K^+\rightarrow \pi^+\pi^0e^+e^-$ with a branching fraction of $4\times10^{-6}$. Other multi-lepton decay modes occur at the level of $10^{-8}$. These are sufficiently small that the background should be negligible with the requirement of $\Ndv\ge2$. If some kaon backgrounds persist in the signal region after these selections, they can be efficiently removed with a cut on the vertex track invariant mass of 0.5 GeV. In Fig.~\ref{fig:trackmass_acceptance}, we show the impact on the acceptance of such a track-mass selection when the LLP decays semi-visibly, finding that the signal acceptance is largely unchanged except at small LLP masses where the typical visible invariant mass is below the value of the cut. The effect of a track-mass cut is even less pronounced for fully visible LLP decays, since the invariant mass of the charged particles in the final state is larger.

\paragraph{Neutral kaons.} The $CP$-even $K_S^0$ state has $c\tau\approx 3$ cm, which is right in the middle of the signal region. $K_S^0$ is therefore potentially the most dangerous background. It typically decays to two charged particles, and so a requirement of $\Ntr\ge3$ greatly suppresses this background. The dominant multi-track decay mode is $K_S^0\rightarrow \pi^+\pi^-e^+e^-$ with a branching fraction of $5\times10^{-5}$. This decay is, however, fully reconstructible, and even a veto inefficiency of 10\% for this decay would still render the background negligible with a requirement of $\Ndv\ge2$. An even more efficient veto, or a veto of vertices with this particular particle combination in the final state, could even make this background manageable with $\Ndv\ge1$ as the requirement. However, this depends on the detailed rate of track mis-reconstruction in $K_S^0$ decays.

The $CP$-odd $K_L^0$ state has $c\tau\approx 1.5$ m. Only about 1\% of $K_L^0$ decays will occur in the signal region for $\mathcal{O}(1)$ boosts. The most dangerous decay mode is $K_L^0\rightarrow \pi^+e^-e^+e^-\nu_e$ with branching fraction $10^{-5}$. Other high-track-multiplicity modes, such as $K_L^0\rightarrow \pi^+\pi^-e^+e^-$, have branching fractions $<10^{-6}$ and are fully reconstructible, and so a veto on the $K_L^0$ peak should efficiently reject them. Together with the lower acceptance of $K_L^0$ decays, the requirement of $\Ndv\ge2$ should suppress the $K_L^0$ backgrounds to negligible level. As with $K^+$ decays, however, if some backgrounds remain then a cut on the vertex track invariant mass of 0.5 GeV should eliminate the remaining kaons in the signal region. As we discussed above and show in Fig.~\ref{fig:trackmass_acceptance}, this cut should not significantly affect the acceptance for most signal benchmarks.

\paragraph{Strange baryons.} Baryon production rates are subdominant to the meson production rates. For example, the off-resonance inclusive $\Lambda$ cross section is about 0.3 pb at $\sqrt{s}=10.52$ GeV \cite{Niiyama:2017wpp}, which is smaller than the total continuum quark cross section of 2.3 pb by almost an order of magnitude. Similarly, the inclusive $B$ meson branching fraction to $\Lambda_c$ baryons (of which 40\% decay to $\Lambda$ baryons) is at the percent level. However, with $c\tau_\Lambda\approx 8$ cm, $\Lambda$ baryons still constitute a major background. Since $\Lambda$ always decay to at most two tracks, requiring $\Ntr\ge3$ should greatly suppress this background. Additionally, the dominant 2-track decay $\Lambda\rightarrow p\pi^-$ is fully reconstructible, and these vertices can be efficiently rejected. The dominant non-fully-reconstructible 2-track decay is $\Lambda\rightarrow pe^-\bar\nu_e$ with a branching fraction of $8\times10^{-4}$. A more targeted search with particle identification selections on the vertex tracks could render a $\Ndv\ge 2$, $\Ntr\ge2$ track search background-free with respect to $\Lambda$ decays.  

Other long-lived strange baryons include the $\Sigma^\pm$ with $c\tau_{\Sigma^\pm} \approx 2.4$ cm, $\Sigma^-$ with $c\tau_{\Sigma^-}\approx4.5$ cm, $\Xi^-$ with $c\tau_{\Xi^-}\approx 4.8$ cm,  $\Xi^0$ with $c\tau_{\Xi^0}\approx8.7$ cm, and $\Omega^-$ with $c\tau_{\Omega^-}\approx2.5$ cm. Potentially problematic decays include $\Sigma^+\rightarrow \Lambda e^+\nu_e,\Lambda\rightarrow p\pi^-$ with branching fraction $2\times10^{-5}$,  $\Xi^-\rightarrow \Sigma^0 e^-\bar\nu_e,\Sigma^0\rightarrow \Lambda\pi^0$ with branching fraction $9\times10^{-5}$, and $\Omega^-\rightarrow \Xi^-\pi^+\pi^-,\Xi^-\rightarrow\Lambda\pi^-$ with branching fraction $4\times10^{-4}$, although the tracks from the subsequent $\Lambda$ or $\Xi^-$ decay should be sufficiently displaced from the original decay vertex that such vertices can be rejected. We therefore conclude that strange baryons can likely be eliminated as a source of background.

\section{Discussion and Conclusions} \label{sec:conclusions}

We have proposed several strategies for inclusive, long-lived particle searches at the Belle II experiment that should greatly expand sensitivity to GeV-scale hidden sectors. To motivate such new searches and characterize their reach, we have developed an EFT framework to classify LLP decay modes in a gauge-invariant manner, finding that multi-track displaced vertex analyses can have sensitivity to a wide range of models. We have also considered multiple well-motivated LLP production modes, including dark photon and dark Higgs portals, finding good sensitivity to couplings of interest from dark matter and muon $g-2$ models. Crucially, many of these extended hidden-sector models would not have been discovered in existing searches, which tend to target specific exclusive production and decay modes.

While we have focused on the Belle II experiment, whose substantial integrated luminosity and clean environment should allow for the low-background reconstruction of multi-track displaced vertices, the hidden-sector models we presented here should also lead to large production rates at LHCb. Because of the large boosts characteristic of forward experiments at hadron colliders, LLP lifetimes must be shorter than at Belle II to give displaced decays in the tracking system. This gives a relatively large increase in $B$ backgrounds, making inclusive searches somewhat more challenging; indeed, most inclusive LLP searches at LHCb only have sensitivity for LLP masses above approximately 20 GeV \cite{Aaij:2016xmb,Aaij:2017mic} because the large vertex mass can be used to efficiently reject $B$ backgrounds, although the reconstruction of displaced resonances is possible at lower masses \cite{Aaij:2020ikh}. While it is possible for LHCb to improve on low-mass exclusive and inclusive signatures \cite{Shuve:2016muy,CidVidal:2019urm} and this merits further study, we nevertheless expect Belle II to have a unique role in covering intermediate-lifetime hidden sectors in an inclusive manner.

\section*{Acknowledgments}
We are grateful to Victoria Lloyd for collaboration at an early stage of this project.
We thank Torben Ferber and Abi Soffer for helpful discussions of LLPs at Belle II,
and Olivier Mattelaer and Qiang Li 
for useful correspondence regarding \madgraph\ and the ISR plugin. 
This manuscript has been authored by Fermi Research Alliance, LLC under 
Contract No. DE-AC02-07CH11359 with the U.S. Department of Energy, Office of Science, Office of High Energy Physics.  This work is supported by the U.S. National Science Foundation under Grant PHY-1820770.

\appendix
\section{Effective Field Theory Framework of LLP Decays}
\label{sec:llpeft}
In this appendix we describe in more detail how we classify LLP decay operators 
in an electroweak-gauge-invariant EFT. This task is made simple by the fact that 
dark sector LLPs are neutral under SM gauge symmetries, which allows us to 
repurpose established results from SMEFT (the collection of higher-dimensional operators 
constructed from SM fields alone)~\cite{Grzadkowski:2010es,Alonso:2014zka,Lehman:2014jma,Passarino:2016pzb} 
and SM + sterile neutrino EFT ($\nu$SMEFT)~\cite{delAguila:2008ir,Bhattacharya:2015vja,Liao:2016qyd}. 
The references provided here contain comprehensive lists of operators which we do not reproduce; instead 
we describe their general structure at different operator dimensions and provide explicit examples 
that give rise to the specific decays studied in the main part of the paper.  

\subsection{Spin-0 LLP}
If the LLP $\chi$ is a scalar, then the decay operator has the generic form $\chi \OSM$ where 
$\OSM$ is also a Lorentz scalar and therefore belongs to SMEFT. From Eq.~\eqref{eq:osm_dimension} we see that we want to focus on operators with $\dim \OSM \leq 7$.

At dimension 4 (corresponding to $\dim \chi \OSM = 5$), we can multiply any term in the 
renormalizable SM Lagrangian  by $\chi$:~the total derivative terms $\widetilde{G}^a_{\mu\nu} G^{a\,\mu\nu}$, $\widetilde{B}_{\mu\nu} B^{\mu\nu}$
and $\widetilde{W}^a_{\mu\nu} W^{a\,\mu\nu}$, and pseudo-scalar couplings to fermions. 
These lead to $\phi \rightarrow \bar f_i f_i$ (for all kinematically accessible fermions other than $\nu$) and $\phi \rightarrow \gamma\gamma, \; g g$, 
with other decays (e.g. to EW gauge bosons) being kinematically forbidden for LLPs of interest to $B$-factories; these final states are shown in the first row of 
Tab.~\ref{tab:s0_final_state_summary}.

At dimension 5 (corresponding to $\dim \chi \OSM = 6$) there is only a single operator $\chi (L_i H)(L_j H)$, which leads to $\phi \rightarrow \nu_i \nu_j$ at low energies after EW symmetry-breaking. 
This decay is shown in the second line of Tab.~\ref{tab:s0_final_state_summary}.

At dimension 6 (corresponding to $\dim \chi \OSM = 7$), there are 59 baryon number $B$-preserving and 4 $B$-violating operators~\cite{Grzadkowski:2010es,Passarino:2016pzb}. 
These interactions are listed in Tab.~1 of Ref.~\cite{Passarino:2016pzb} and Eq.~(1) of Ref.~\cite{Alonso:2014zka}, respectively. The LLP decay final states enabled 
by these operators are listed in the third line of Tab.~\ref{tab:s0_final_state_summary}. The leptonic, semi-leptonic and hadronic decay operators for scalar 
LLPs used in the body of the paper are generated by the following operators at this dimension:
\beq
\chi (\bar \ell_p \gamma_\mu \ell_r)(\bar \ell_s \gamma^\mu \ell_t),\;\; 
\chi (\bar L_p^j \ell_r) \epsilon_{jk} (\bar Q_s^k u_t), \;\;
\chi (\bar u_p \gamma_\mu u_r)(\bar u_s \gamma^\mu u_t).
\label{eq:ew_symmetric_scalar_ops}
\eeq
where lower (upper) case fields are right- (left-) handed Weyl fermions, and the Roman indices are flavour labels.

All  dimension-7 operators (corresponding to $\dim \chi \OSM = 8$) violate $B-L$, and some break $B$ or $L$ individually~\cite{Lehman:2014jma,Henning:2015alf}, making them 
particularly interesting. However, we leave these for a future study as symmetry-violating operators deserve a careful treatment as they are
likely to be strongly constrained by other observables. 


\subsection{Spin-1/2 LLP}
Fermionic LLPs $\chi$ are SM-neutral fermions and therefore identical to sterile neutrinos. 
The decay operators $\chi \OSM$ then directly correspond terms in $\nu$SMEFT~\cite{delAguila:2008ir,Bhattacharya:2015vja,Liao:2016qyd}.
Using Eq.~\eqref{eq:osm_dimension}, we now need to enumerate operators with  
dimension $\leq 13/2$, that is, at least one SM fermion and a dimension-5 object.
At $\dim \chi \OSM = 4$ the only possibility is $ L_i H \chi$, which gives rise 
to $\chi-\nu_L$ mass-mixing, allowing $\chi$ to decay via mixing as a right-handed neutrino \cite{Shrock:1980ct,Shrock:1981wq,Gronau:1984ct}.  
There are no $\dim \chi \OSM = 5$ operators with a single $\chi$~\cite{Aparici:2009fh,Alonso:2014zka}.

The $\dim \chi \OSM = 6$ operators are listed in Tab.~1 of Ref.~\cite{Liao:2016qyd}. We use 
\beq
(\bar{d}_p\gamma^\mu u_r)(\bar{\chi}\gamma_\mu \ell_s)
\eeq
to generate semileptonic decays of a spin-$1/2$ LLP as described in the main text. We show this 
and other possible final states produced at this dimension in the first line of Tab.~\ref{tab:s12_final_state_summary}.

Many operators at $\dim \chi \OSM = 7$ (collected in Eqs.~(14)-(17), (22) and (25) of Ref.~\cite{Liao:2016qyd}) are obtained by adding covariant derivative to a 
dimension-6-like operator. Thus these operators generate the same final states as dimension six, possibly with an addition gauge boson 
(with $\gamma$ and $g$ being the most relevant in our mass range). The corresponding decay channels 
are listed in the second line of Tab.~\ref{tab:s12_final_state_summary}.

\section{Hadronic Final States}
\label{sec:hadronization}

In considering LLP decays to hadronic final states, we need to  translate parton-level operators for LLP decays
into operators involving mesons and baryons. For large LLP masses, it is appropriate to decay the LLP to quarks and then rely
on a parton shower/hadronization model to obtain the hadronic final state. Conversely, for small LLP masses, chiral perturbation
theory (ChiPT) and the Hidden Local Symmetry (HLS) frameworks provide a way of estimating the mapping of quark operators to meson
operators. The boundary between these two treatments is not clearly defined, although we expect that the parton model
should give reasonable results for $q^2\gtrsim\mathrm{GeV}^2$, where $q$ is the 4-momentum flowing into the composite quark operator. 
For each operator involving quarks, we perform calculations with both  chiral perturbation theory and a parton shower + hadronization,
and compare the results in the region of intermediate hadronic momentum where neither treatment is fully valid. We find
that in all cases we do get charged track multiplicities of three or more at appreciable rates, although in certain regimes the two
treatments differ in the degree to which this is true. 

We emphasize that, for the purposes
of the experimental search, the theoretical uncertainty on the hadronization model is irrelevant:~multi-track searches are well
motivated by both approaches to hadronization, and as long as the search results are transparent in terms of the precise model used in 
setting limits (or with efficiencies provided for displaced object reconstruction), the results can be reinterpreted and applied to 
other, more refined hadronization models, particularly as  theories of hadronic GeV-scale decays continue to be refined.

\subsection{\texorpdfstring{$\bar u_{R}\gamma^\mu d_{R}$}{uR dR} Operator} \label{sec:vector_ud_operator}
This operator closely resembles the hadronic current in leptonic meson decays. For example, 
for exclusive decays into hadrons we can use  the hadronic matrix elements
\begin{eqnarray}
\langle 0|\bar u\gamma^\mu\gamma^5 d|\pi^-(k)\rangle &=& if_\pi k^\mu, \label{eq:matelem_1}\\
\langle 0|\bar u\gamma^\mu d|\rho^-(k)\rangle &=& \sqrt{2}g_\rho\,\epsilon^\mu,\label{eq:matelem_2}
\end{eqnarray}
where $\epsilon^\mu$ is the $\rho$ meson polarization vector, $f_\pi = 0.130\,\,\mathrm{GeV}$, and $g_\rho = 0.119\,\,\mathrm{GeV}^2$ \cite{Donoghue:1992dd}. For the
operator $ \frac{1}{\Lambda^2} \left(\bar u_{R}\gamma_\mu d_{R} \right)\left( \ell_{R} \gamma^\mu \chi\right)$, we can then calculate the decay rates for 
$\chi\rightarrow e^- \pi^+$ and $\chi\rightarrow e^-\rho^+,\,\rho^+\rightarrow\pi^+\pi^0$ (neglecting the pion mass):
\begin{eqnarray}
\Gamma(\chi\rightarrow e^-\pi^+) &=& \frac{|V_{ud}|^2 f_\pi^3 m_\chi^2}{128\pi\,\Lambda^4},\\
\Gamma(\chi\rightarrow e^-\rho^+) &=& \frac{|V_{ud}|^2 g_\rho^2 m_\chi^3}{64\pi\,\Lambda^4 m_\rho^2}\left(1-\frac{m_\rho^2}{m_\chi^2}\right)^2\left(1+\frac{2m_\rho^2}{m_\chi^2}\right),
\end{eqnarray}
where $V_{ij}$ is the Cabibbo-Kobayashi-Maskawa (CKM) matrix. These results are essentially the same as the exclusive semileptonic decay rates of  heavy neutral leptons \cite{Gorbunov:2007ak}. For $m_\chi\gg m_\rho$, we find that the ratio of decays to $\rho^+$ relative to $\pi^+$ approaches two. Thus, while both of these decay modes produce a single charged pion, the majority of LLP decays actually give an additional $\pi^0$, and so an exclusive reconstruction of the LLP mass will cut out most of the signal events. In our analysis, we neglect the decay into $K^+$, which is both Cabibbo- and phase-space-suppressed.

Higher charged pion multiplicities can occur once we take into account heavier vector mesons. It is well established that the 3-pion $\tau$ decays predominantly arise from $a_1$ intermediate states \cite{Isgur:1988vm}. Similarly, the decay $\chi\rightarrow e^-{a_1^+}^{(*)},\, {a_1^+}^{(*)}\rightarrow \rho^0\pi^+,\,\rho^0\rightarrow \pi^+\pi^-$ gives rise to a four-track decay $\chi\rightarrow e^-\pi^+\pi^-\pi^+$. The calculation of this process is complicated by the large, energy-dependent $a_1^+$ width, as well as nearby broad resonances such as the $a_1'$. There are several models of 3-pion $\tau$ decays. We choose the model of Ref.~\cite{Vojik:2010ua} due to its simplicity and the readiness with which it can be implemented in the \texttt{UFO} format. There are some aspects of the model that cannot be simply accounted for in \madgraph (for example, the renormalization group running of the $a_1$ mass and the energy-dependent $a_1$ width); however, we have found that our somewhat crude implementation of the model predicts the $\tau^-\rightarrow \pi^-\pi^+\pi^-\nu_\tau$ decay width in agreement with the observed value within 15\%, and that the kinematic acceptance for our proposed displaced vertex search in Sec.~\ref{sec:analysis} is insensitive to model details. Away from the $\tau$ mass, the variation in the 3-pion decay rate is of $\mathcal{O}(1)$ when we change model parameters. 

We implement an exclusive hadronic decay model in the \texttt{UFO} format and simulate events using \madgraph\ for $m_\chi < 2$ GeV. At higher masses, we directly implement the decay $\chi\rightarrow e^- u\bar{d}$ in \madgraph, and perform showering and hadronization with  \pythia. In the region $1.5\,\,\mathrm{GeV}<m_\chi<2\,\,\mathrm{GeV}$, we use both methods and compare their acceptance and efficiency in our analysis.

\subsubsection*{Meson Model Details \& Validation}

Our model consists of the LLP, $\chi$, as well as the $\rho^\pm$, $\rho^0$, $\pi^\pm$, $\pi^0$, $a_1^\pm$, and ${a_1^\pm}'$ mesons, and is based on Ref.~\cite{Vojik:2010ua}. The charged pions and vector mesons are coupled to $\chi$ by replacing the hadronic currents with the meson fields to give the matrix elements in Eqs.~\eqref{eq:matelem_1}-\eqref{eq:matelem_2}. The $a_1$ hadronic matrix element is the same as for the $\rho$, but using $g_a=0.250\,\,\mathrm{GeV}^2$ as the decay constant.

The $a_1$ interaction Lagrangian is taken to be
\begin{eqnarray}
\mathcal{L}_{a_1} &=& \frac{g_{a_1\rho\pi}}{\sqrt 2}\left(\mathcal{L}_1\cos\theta+\mathcal{L}_2\sin\theta\right),
\end{eqnarray}
where $\sin\theta\approx 0.463$ and
\begin{eqnarray}
\mathcal{L}_1 
&=& {A^+}^\mu\left(\partial^\nu\pi^-\rho_{\mu\nu}^0-\partial^\nu\pi^0\rho^-_{\mu\nu}\right)+{A^0}^\mu\left(\partial^\nu\pi^+\rho_{\mu\nu}^--\partial^\nu\pi^-\rho_{\mu\nu}^+\right) + {A^-}^\mu\left(\partial^\nu\pi^0\rho^+_{\mu\nu}-\partial^\nu\pi^+\rho_{\mu\nu}^0\right),\nonumber\\
\mathcal{L}_2 
&=& \rho_{\mu\nu}^+\left(\pi^-\partial^\mu {A^0}^\nu-\pi^0\partial^\mu {A^-}^\nu\right) + \rho_{\mu\nu}^0\left(\pi^+\partial^\mu{A^-}^\nu - \pi^-\partial^\mu{A^+}^\nu\right)+\rho^-_{\mu\nu}\left(\pi^0\partial^\mu {A^+}^\nu-\pi^+\partial^\mu{A^0}^\nu\right)\nonumber
\end{eqnarray}
with $A^{0,\pm}_\mu$ being the field corresponding to $a_1$.
The coupling $g_{a_1\rho\pi}$ is fixed by obtaining the observed $a_1$ width, and we obtain $g_{a_1\rho\pi} = 54\,\,\mathrm{GeV}^{-1}$. 

The Lagrangian describing the $\rho$ coupling to pions is
\begin{eqnarray}
\mathcal{L}_\rho &=& ig_{\rho\pi\pi}\left({\rho^0}^\mu\pi^-\overset{\leftrightarrow}\partial_\mu\pi^++{\rho^+}^\mu\pi^0\overset{\leftrightarrow}\partial_\mu\pi^-+{\rho^-}^\mu\pi^+\overset{\leftrightarrow}\partial_\mu\pi^0\right),
\end{eqnarray}
where $g_{\rho\pi\pi} \approx 6$ gives the correct $\rho$ width \cite{Klingl:1996by}. 
The authors of Ref.~\cite{Vojik:2010ua} found that the inclusion of the $a_1'$, or $a_1(1640)$, is important to get good agreement with 3-pion $\tau$ decays. The hadronic matrix element  is the same for the $a_1'$ as for the $a_1$, but with an additional multiplicative complex constant $\alpha = -0.31\pm0.32i$. The $a_1'$ is taken to have the same hadronic couplings as the $a_1$. In our model, we use $m_{a_1} =1.232\,\,\mathrm{GeV}$, $\Gamma_{a_1}=0.431\,\,\mathrm{GeV}$, $m_{a_1'}=1.655\,\,\mathrm{GeV}$, and $\Gamma_{a_1'}=0.254\,\,\mathrm{GeV}$.

The model of Ref.~\cite{Vojik:2010ua} implements a form-factor suppression of fully hadronic vertices; however, for the kinematic range of interest to us, this is not a significant effect and would greatly add to the complexity of the model and so we neglect it. The model of Ref.~\cite{Vojik:2010ua} also includes the $\sigma$, or $f(500)$, scalar meson. Its contribution to the 3-pion decays of the $a_1$ meson are only at the 20\% level, and as the $\sigma$ meson is extremely broad (its width is comparable to its mass), it is not straightforward to model. Therefore, we do not include the $\sigma$ meson in our model.

We implemented this model in the \texttt{UFO} format using \texttt{FeynRules}, and we use \madgraph\ to simulate events. To validate the model, we simulated the process $\chi\rightarrow \pi^+\pi^-\pi^+e^-$ for $m_\chi = m_\tau$ and compared the results to CLEO data of $\tau^-\rightarrow \pi^-\pi^+\pi^-\nu_\tau$~\cite{Asner:1999kj,Vojik:2010ua}. Our model predicts a value for the 3-pion partial width that is 87\% of the PDG 3-pion width of the $\tau$; this is remarkably good agreement, and the discrepancy is at the level of the $\sigma$ meson contribution that we neglect in our model. We can then calculate the $\chi\rightarrow \pi^+\pi^-\pi^+e^-$ partial width at other values of the $\chi$ mass to determine the branching fraction to 4 charged tracks, applying the same correction factor of $1/0.87$ to account for decay modes not included in our model. 

    \begin{figure}[t]
      \centering
      \includegraphics[width=0.4\textwidth]{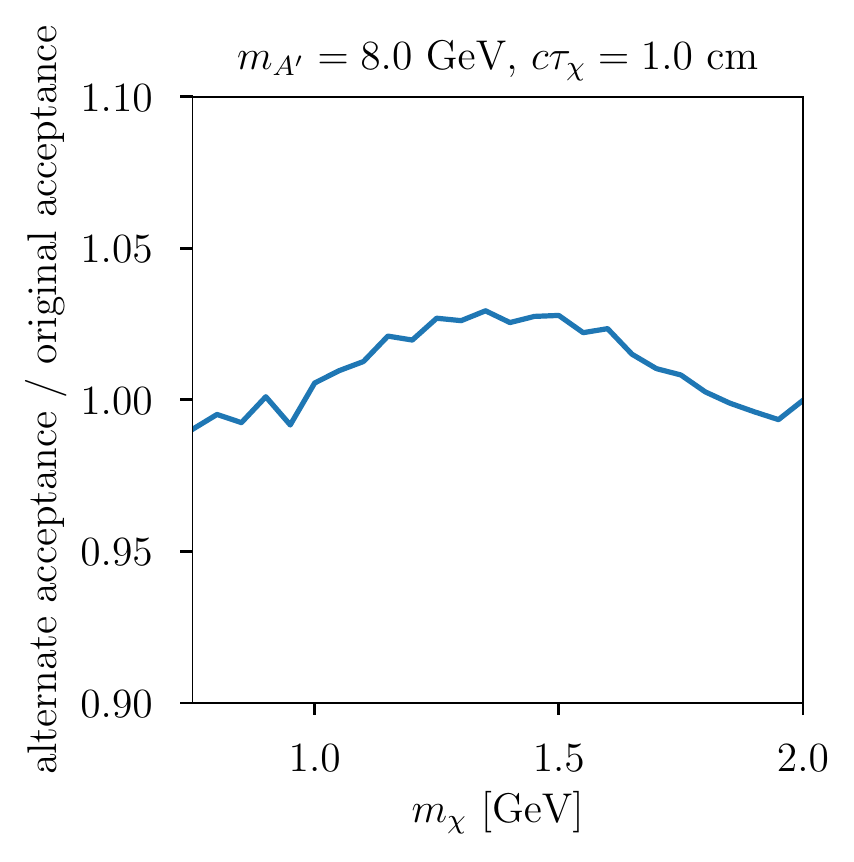}
      \caption{
       Ratio of acceptances of a $N_{\rm dv}\ge1$, $N_{\rm tr}\ge3$ search with a $m_{A'}=8\,\,\mathrm{GeV}$, $c\tau_\chi=1\,\,\mathrm{cm}$ signal for the original $\chi\rightarrow \pi^+\pi^-\pi^+e^-$ decay model compared to the alternate model where we have adjusted some of the intermediate resonance masses. This ratio is within 5\% of unity over the mass range for which we use the exclusive meson decay model.
      \label{fig:acceptance_darkphoton_fudgedcompare_mAp8GeV_ctau1cm}}
    \end{figure}

We also used our \madgraph\ model to compute the $m_{\pi\pi\pi}$ invariant mass  spectrum from  $\chi\rightarrow \pi^+\pi^-\pi^+e^-$ and compare with CLEO data from $\tau^-\rightarrow \pi^-\pi^+\pi^-\nu_\tau$ decays. Our spectrum is slightly broader than theirs and with the peak shifted 0.2 GeV to the right. We therefore found an alternative model benchmark point where we obtained good agreement with the data if we shift the masses to $m_{a_1}=1$ GeV and $m_{a_1'}=1.35$ GeV. Clearly this is unphysical, but as a phenomenological model it allowed us to obtain a 3-pion spectrum that is consistent with the data from $\tau$ decays and we can use this to estimate a ``systematic'' on the signal acceptance by varying these parameters. We  repeated our calculation of the kinematic and geometric acceptance of a search for a single displaced vertex with at least three tracks on the sample with 3-pion decays; we show the ratio of acceptances for the two models in Fig.~\ref{fig:acceptance_darkphoton_fudgedcompare_mAp8GeV_ctau1cm}. There is no appreciable difference in the acceptance of the two models, which gives us confidence  in the robustness of our results in spite of the \emph{ad hoc} nature of some aspects of our hadronic model. 

We can also use our alternative model benchmark to calculate the 3-pion partial width of $\chi$, normalizing to the $\tau$-mass value as before. We find that the two models agree for $m_\chi\gtrsim1.5$ GeV, but unsurprisingly the alternative model gives a larger 3-pion width at lower mass due to the lower $a_1/a_1'$ masses, which allows for on-shell contributions for a wider range of $\chi$ masses. The discrepancy between the two models can be as much as a factor of 3, although we expect the original model to give a better estimate of the 3-pion partial width. This does suggest that the uncertainty on the 3-pion width from our method is of $\mathcal{O}(1)$, but given that we see extremely good sensitivity of our multi-track vertex search, this does not alter the motivation for the search or our general conclusions.

\begin{figure}
      \centering
      \includegraphics[width=0.7\textwidth]{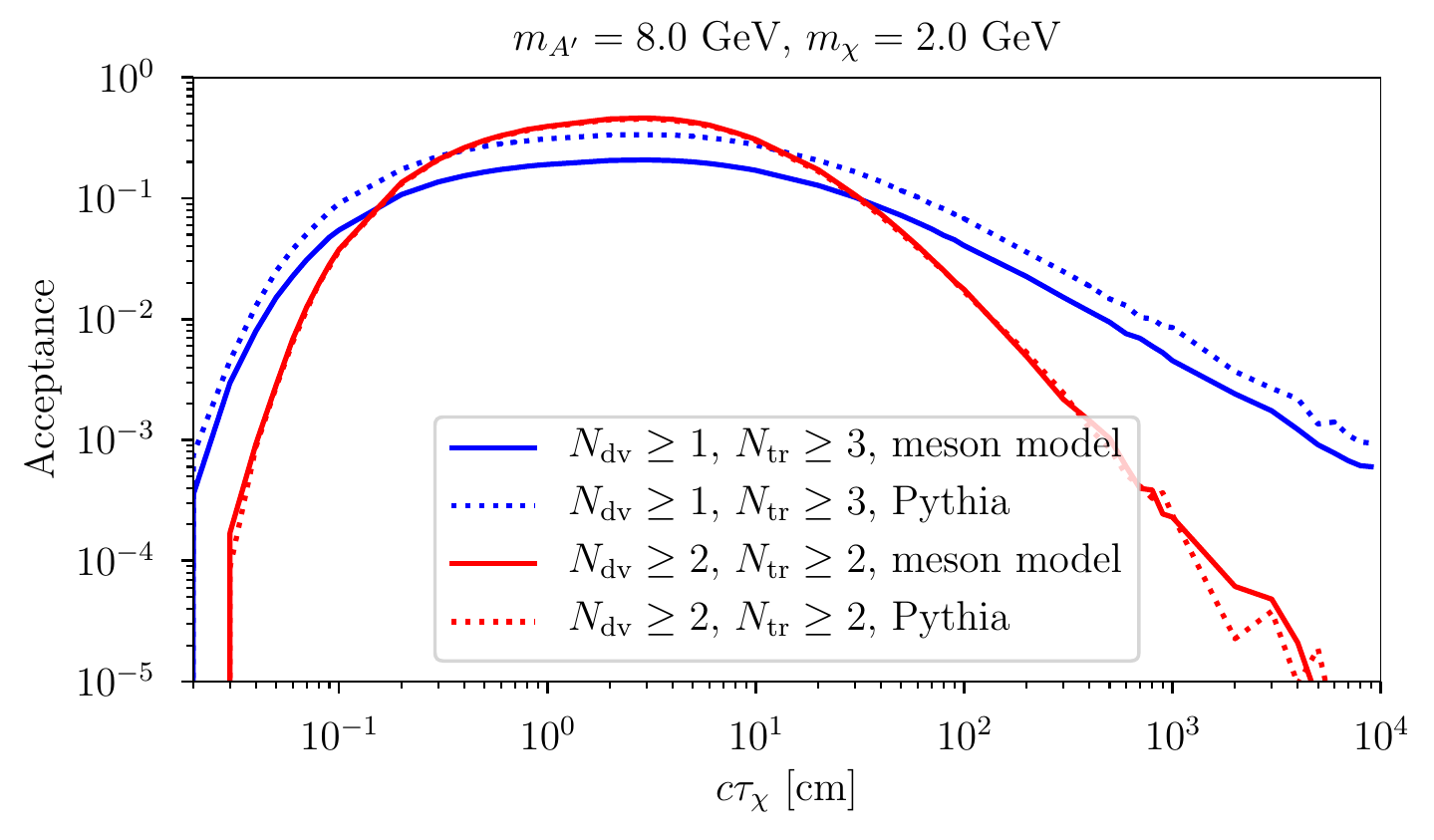}
      \caption{Acceptance of searches from Sec.~\ref{sec:analysis} for $\chi\to e\bar{d}u$, comparing results from the meson model of Appendix~\ref{sec:vector_ud_operator} with \pythia. We consider a benchmark of $m_{A'}=8\;\GeV$, $m_\chi=2\;\GeV$. \label{fig:acceptance_slices_edu_pythia_vs_chipt}}
    \end{figure}

Finally, we compute the acceptances of our proposed displaced vertex searches from Sec.~\ref{sec:analysis}  twice:~using the meson decay model, as well as with decays of $\chi$ to quarks using \pythia\ for showering and hadronization. We expect that \pythia\ should give the more correct description at larger LLP masses, while exclusive meson decays provide a better description for small LLP masses. We compare the acceptances of the two approaches for $m_\chi=2\;\GeV$ in Fig.~\ref{fig:acceptance_slices_edu_pythia_vs_chipt}, finding that the two descriptions agree very well for the $\Ndv=2$, $\Ntr=2$ analysis, and agree within a factor of 2 for the $\Ndv=1$, $\Ntr=3$ analysis (the acceptance of  the $\Ndv =2$, $\Ntr=3$ analysis is approximately the square of the $\Ndv=1$, $\Ntr=3$ analysis). Given the challenge of correctly modelling hadronization in this intermediate mass range, this gives us confidence in the robustness of our results. We find a similar comparison between the two hadronization methods throughout the mass range 1.5--2 GeV.

    \subsection{\texorpdfstring{$\bar d_{L} u_{R}$}{dL uR} Operator} \label{sec:scalar_du_operator}
The mapping of quark-level operators into ChiPT is discussed in, \emph{e.g.,} Ref.~\cite{Claudson:1981gh} in
the context of GUT-scale $B$-violating interactions leading to proton decay. First, we need to identify 
the chiral transformation properties of the operator
\beq
O_{lequ} = \chi (\bar \nu_L \ell) (\bar q_L C_{lequ} q_R),
\label{eq:quark_level_scalar_semileptonic_op}
\eeq
whose EW-symmetric form is given in Eq.~\eqref{eq:ew_symmetric_scalar_ops} and the quarks $q_{L(R)}$ transform as a $3$ of $SU(3)_{L(R)}$; 
we also introduced a $3\times 3$ flavour-space matrix $C_{lequ}$. 
The choice 
\beq
C_{lequ} =  \begin{pmatrix}
  0 & 0 & 0 \\
  1 & 0 & 0 \\
  0 & 0 & 0 
\end{pmatrix}
\eeq
generates $\chi(\bar\nu_L e_R)( \bar{d}_L u_R)$ used in the main text. 
We treat $C_{lequ}$ as a spurion of the chiral $SU(3)_L\times SU(3)_R$ symmetry, which 
transforms as 
\beq
C_{lequ}\rightarrow L C_{lequ} R^\dagger,
\label{eq:spurion_tform}
\eeq
where $L$ ($R$) are elements of $SU(3)_{L\,(R)}$. 
This transformation rule makes it easy to identify the leading operator 
corresponding to $O_{lequ}$ in chiral perturbation theory, since $C_{lequ}$ 
transforms in the same way as the quark mass matrix. Therefore $O_{lequ}$ maps onto 
\begin{align}
  O_{lequ} & \to  B\chi (\bar\nu_L e_R)\tr\left[\Sigma^\dagger C_{ledu}\right] + \hc \supset + \frac{i\sqrt{2}B}{f_\pi} (\bar\nu_L e_R)\pi^+ + \hc+\dots
\label{eq:lequ_in_chipt}
\end{align}
where  $\Sigma = \exp(2 i \Pi /f_\pi)$ and $\Pi$ is the meson field\footnote{We only consider the octet $\eta$ meson for simplicity; the effects 
of mixing with the singlet $\eta'$ can be included as described in, \emph{e.g.,} Ref.~\cite{Donoghue:1992dd}.} 
\beq
\Pi =\frac{1}{2} \begin{pmatrix}
  \pi_0 + \frac{1}{\sqrt{3}}\eta & \sqrt{2}\pi^+ & \sqrt{2} K^+ \\
  \sqrt{2}\pi^- & \frac{1}{\sqrt{3}}\eta - \pi_0 & \sqrt{2} K_0 \\
  \sqrt{2} K^- & \sqrt{2}\bar{K} & -\frac{2}{\sqrt{3}} \eta
\end{pmatrix}.
\eeq
The exponentiated field transforms as
\beq
\Sigma \rightarrow L\Sigma R^\dagger.
\eeq
The Wilson coefficient $B$ can be estimated by analogy with the chiral mass term which has the same structure 
as the quark part of $O_{lequ}$ and gives rise to meson masses; this comparison yields $B \sim m_\pi^2 f_\pi^2/(2(m_u+m_d)) \sim (300\;\MeV)^3$.

In the last step of Eq.~\eqref{eq:lequ_in_chipt} we only showed the leading term in $1/f_\pi$; higher-order terms 
contain more mesons, and at $\mathcal{O}(1/f_\pi^3)$ terms with multiple charged mesons appear. 

We are interested in the decays of LLPs with mass of $\mathcal{O}(\GeV)$ which is close to the cut-off of ChiPT of $\sim 4\pi f_\pi$, implying the possible importance of vector mesons in multi-track decays. 
We construct operators containing vector mesons next.

\subsubsection*{Including Vector Mesons}

Multiple charged tracks may be generated through the production and 
decay of vector mesons. We implement vector meson couplings using the Hidden Local Symmetry (HLS) formalism~\cite{Harada:2003jx}, 
in which the building blocks are $\xi_{L,R}$:
\beq
\xi_{L,R} = \exp\left(\mp i \Pi/f_\pi\right).
\eeq
These fields are related to usual ChiPT field $\Sigma$ via
\beq
\Sigma = \xi_L^\dagger \xi_R.
\eeq
Unlike $\Sigma$, however, they transform non-trivially under the entire (global + hidden) symmetry group:
\beq
\xi_{L,R} \to h(x) \xi_{L,R} (L^\dagger,R^\dagger),
\eeq
where $h(x)\in SU(3)_V$ is an element of the hidden gauge group, and $L\;(R)\in SU(3)_{L\,(R)}$ as before.
In order to find an operator that can lead to vector meson production, we need to use a modified spurion analysis which ensures that the result is invariant under the hidden gauge symmetry, and it is not degenerate with using $\Sigma$ (i.e., does not just reduce to Eq.~\eqref{eq:lequ_in_chipt}).
One way to achieve this is to start from the (hidden) gauge-covariant derivatives
\beq
D_\mu \xi_{L,R} = \partial_\mu \xi_{L,R} - i g_V V_\mu \xi_{L,R},
\eeq
where $g_V \approx 5.9$ is the hidden gauge coupling, and $V_\mu$ is the vector meson matrix
\beq
V = \frac{1}{2}\begin{pmatrix}
\omega_0 + \rho_0 & \sqrt{2}\rho^+ & \sqrt{2}K^{*+} \\
\sqrt{2} \rho^- & \omega_0 - \rho_0 & \sqrt{2}K^{*0} \\
\sqrt{2} K^{*-} & \sqrt{2}\bar{K}^{*0} & \sqrt{2}\phi 
\end{pmatrix}.
\eeq
The covariant derivatives of $\xi$ have the usual simple transformation property under the hidden gauge symmetry:
\beq
D_\mu \xi_{L,R} \to h(x) D_\mu \xi_{L,R}. 
\eeq
The simplest operator that preserves HLS and the global chiral symmetry is 
\begin{align}
  O_{lequ}^{\mathrm{HLS}} & =  f_\pi \chi (\bar\nu_L e_R)\tr\left[\left(D_\mu \xi_L \right) C_{lequ} \left(D_\mu\xi_R\right)^{\dagger}\right] + \hc \label{eq:lequ_in_hls} \\
  & \supset \frac{g_V}{\sqrt{2}} (\bar\nu_L e_R) \left( \rho^{+,\mu} \partial_\mu \pi^0 - \rho^{0,\mu} \partial_\mu \pi^+\right) + \hc + \dots 
\end{align}
where in the last line we are showing only a single term from the $1/f_\pi$ expansion of $\xi_{L,R}$.
We estimated the Wilson coefficient by considering the analogous term without the leptonic factor, which contains 
vector meson masses $m_V^2 \sim g_{V}^2 f_\pi^2 $ if $C_{lequ}  = 1$.
Since the decays $\rho^+ \to \pi^+ \pi^0$ and $\rho^{0} \to \pi^+ \pi^-$ have $\mathcal{O}(1)$ branching fractions, these operators can easily give rise to multi-track 
signals, while avoiding the phase space and chiral ($1/f_\pi$) suppression of the multi-body decays involving only mesons (if the vectors can be produced on shell).

\subsubsection*{Comparison with \pythia}
\begin{figure}
      \centering
      \includegraphics[width=0.47\textwidth]{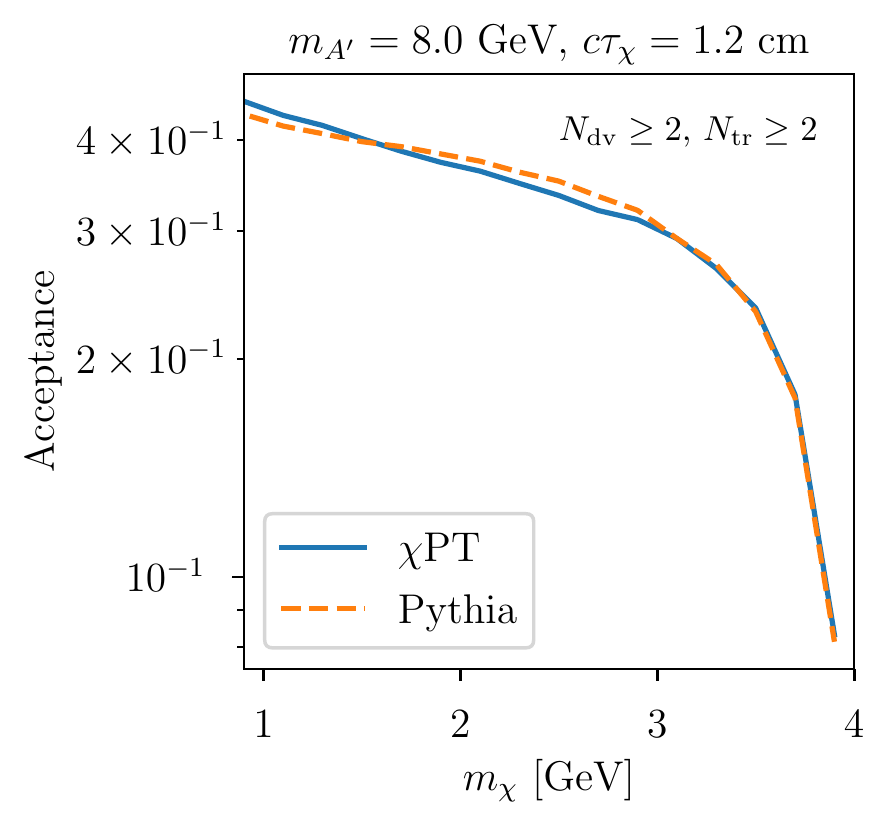}
      \includegraphics[width=0.47\textwidth]{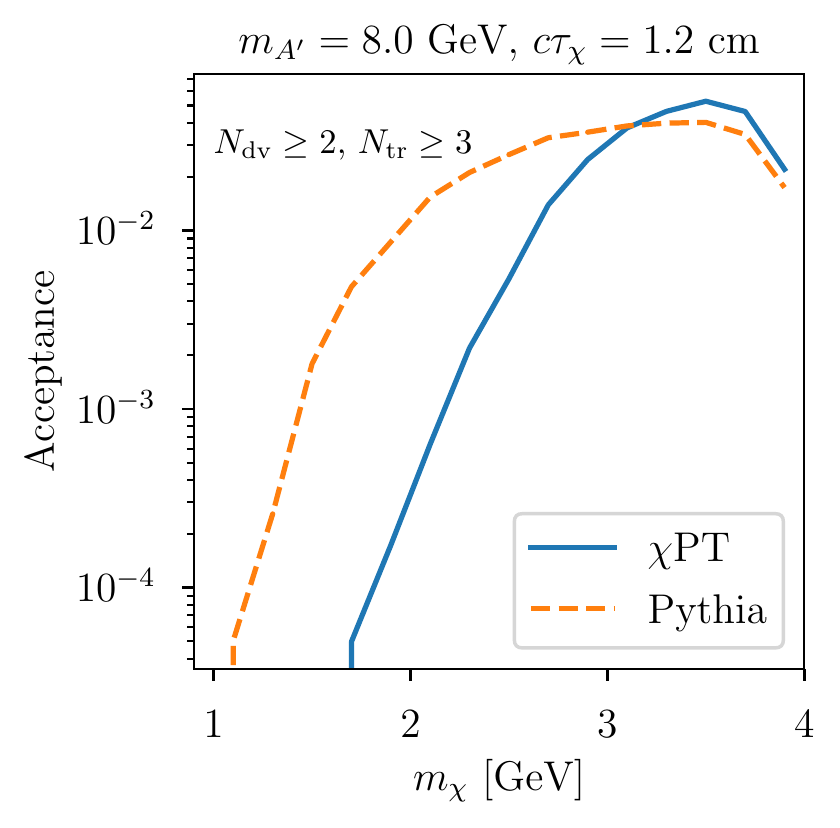}
      \caption{Comparison of LLP signal acceptances evaluated in two different hadronization models of $\chi \to \nu e^+ \bar{u}d$ for $\mAp = 8\;\GeV$ and $c\tau_\chi \approx 1\;\cm$. 
        In the left (right) panel we show the $\Ndv \geq 2$, $\Ntr \geq 2$ ($\Ndv \geq 2$, $\Ntr \geq 3$) analyses. 
        The solid blue lines show acceptances for final state generated using chiral perturbation theory/hidden local symmetry approach, 
        while the orange dashed lines were obtained using \pythia.
      \label{fig:acceptance_slices_nuedu_pythia_vs_chipt}}
    \end{figure}

We have implemented the hadronization of the scalar semileptonic operator in Eq.~\eqref{eq:quark_level_scalar_semileptonic_op} using \pythia, and by incorporating the chiral-perturbation theory and HLS results described above in a \texttt{FeynRules} model. 
In the latter approach we include terms in the $1/f_\pi$ expansion that enable 
$\chi \to \nu e^+ (\pi^-,\;\pi^- \eta,\; \pi^- \pi \pi,\;\pi \rho)$ and allowing 
the heavier mesons to decay as $\eta \to \pi^0 \pi^+ \pi^-$ and $\rho\to \pi\pi$; 
depending on the charges of $\rho$ and $\pi$ these decay channels can produce up to four 
tracks per vertex. Note that there is an ambiguity in the relative branching fractions to final states 
involving vector mesons, since the Wilson coefficients in Eqs.~\eqref{eq:lequ_in_chipt} and~\eqref{eq:lequ_in_hls} 
are only fixed by inexact arguments. We have generated full event samples using both hadronization approaches and 
computed the signal acceptance using the selections discussed in Sec.~\ref{sec:analysis}. 
In Fig.~\ref{fig:acceptance_slices_nuedu_pythia_vs_chipt}, we compare acceptances for a single point in 
the LLP parameter space with $\mAp = 8\;\GeV$, and $c\tau_\chi \approx 1\;\cm$.
In the left panel we show the $\Ndv\geq 2$ and $\Ntr\geq 2$ analysis which illustrates remarkable 
agreement between the two approaches. In this case, the dominant decay channel is $\chi \to \nu e^+ \pi^-$ 
which is already captured by the leading term in the chiral expansion. Significant differences 
emerge in the $\Ndv\geq 2$ and $\Ntr \geq 3$ analysis shown in the right panel of Fig.~\ref{fig:acceptance_slices_nuedu_pythia_vs_chipt}; 
we see that the chiral perturbation/HLS treatment of hadronization leads to a smaller acceptance for $m_\chi \lesssim 3\;\GeV$ compared 
to \pythia. We can understand this discrepancy as follows. In decays mediated by Eq.~\eqref{eq:quark_level_scalar_semileptonic_op}, 
about half of the rest mass energy of $\chi$ flows into the hadronic part of the operator; demanding $\geq 3$ tracks further 
partitions this energy into a number of particles with total mass of at least $3m_\pi$, suggesting that the 
phase space suppression from finite meson masses becomes important. We have confirmed that this an 
important effect by evaluating partial widths to different hadronic final states at unphysically low meson masses 
in the chiral perturbation/HLS approach.\footnote{We have also compared the multi-hadron branching fraction estimated from \pythia, dimensional analysis and ChiPT/HLS, finding that the first two are in good agreement with each other throughout the LLP mass range. This further supports the hypothesis that phase space suppression is not adequate for low LLP masses in the \pythia hadronization model.} Additionally, the invariant mass distribution of the $ud$ pair is broad, and 
even in the $m_\chi \sim 3\;\GeV$ case half of the events have $\sqrt{s_{ud}} \lesssim 1.2\;\GeV$. 
These observations suggest that the \pythia hadronization model most likely overestimates the 
number of high-multiplicity events, leading to a higher signal acceptance for 
$\GeV$-scale LLP masses. 
The chiral perturbation theory/HLS approach, however, is also deficient because we do not know how 
to precisely fix the relative sizes of different operators that contribute to the decay. It is reassuring that both methods agree for larger LLP masses $m_\chi \gtrsim 3\;\GeV$.

    \subsection{\texorpdfstring{$\bar{u}_R u_R \bar{u}_R u_R$}{uR uR uR uR} Operator} \label{sec:scalar_uuuu_operator}
Fully hadronic decays can be modelled in ChiPT/HLS similarly to the semi-leptonic case.
We begin with the operator 
\beq
O_{uuuu} = \chi (\bar q_R C_{uu}\gamma_\mu q_R)(\bar q_R C_{uu} \gamma^\mu q_R),
\label{eq:quark_level_scalar_hadronic_op}
\eeq
whose EW-symmetric form is given in Eq.~\eqref{eq:ew_symmetric_scalar_ops}; $C_{uu}$ is a spurion of 
$SU(3)_{L} \times SU(3)_R$ which must have the expectation value
\beq
C_{uu} =  \begin{pmatrix}
  1 & 0 & 0 \\
  0 & 0 & 0 \\
  0 & 0 & 0 
\end{pmatrix}.
\eeq
to generate the $\chi (\bar{u}_R u_R) (\bar{u}_R u_R)$ operator.
Chiral invariance demands that this matrix transforms as 
\beq
C_{uu} \to R C_{uu} R^\dagger.
\eeq
Since there are no $SU(3)_L$ transformations involved, we 
can build our operators out of objects charged only under $SU(3)_R$, 
such as $\xi_R^\dagger D_\mu \xi_R$ which transforms as 
\beq
\xi_R^\dagger D_\mu \xi_R \to R (\xi_R^\dagger D_\mu \xi_R) R^\dagger
\eeq
These objects are also invariant under the HLS. 
Some invariant combinations are
\begin{subequations}
\begin{align}
  \chi \left( \tr\left[\xi_R^\dagger D_\mu \xi_R C_{uu}\right]\right)^2 & \supset -\frac{1}{4} (\rho^0_\mu)^2 + \frac{1}{2f_\pi} \rho^0_\mu \partial_\mu \pi^0 - \frac{1}{4f_\pi^2} (\partial_\mu \pi^0)^2 + \dots \label{eq:uuuu_hls_1}\\ 
  \chi \tr C_{uu} \tr \left[\xi_R^\dagger D_\mu \xi_R C_{uu}\xi_R^\dagger D_\mu \xi_R\right] & \supset -\frac{1}{4} (\rho^0_\mu)^2 + \frac{1}{2f_\pi} \rho^+_\mu \partial_\mu \pi^- - \frac{1}{2f_\pi^2} (\partial_\mu \pi^+ \partial_\mu \pi^-) + \dots  \label{eq:uuuu_hls_2}\\
  \chi \tr \left[\xi_R^\dagger D_\mu \xi_R C_{uu} \xi_R^\dagger D_\mu \xi_R C_{uu}\right] &  \supset -\frac{1}{4} (\rho^0_\mu)^2 + \frac{1}{2f_\pi} \rho^0_\mu \partial_\mu \pi^0 - \frac{1}{4f_\pi^2} (\partial_\mu \pi^0)^2 + \dots
\end{align}
\end{subequations}
As before, we only highlighted a few representative terms, but there are many more that can be important. Of particular interest to us are those containing 
$\eta$, $\omega_0$ and other mesons that can decay to multiple charged particles. 
Unfortunately we do not know a way to fix the relative sizes of these operators, so the branching fractions of the 
different hadronic states are uncertain. For concreteness, we take the coefficient of the operator leading to $\chi \to \pi^+\pi^-$ to be 
four times larger than the others, since this leads to a better agreement with \pythia in the $\Ntr \geq 2$ analysis as we show below.

\subsubsection*{Comparison with \pythia}
We have implemented the hadronization of the scalar hadronic operator in Eq.~\eqref{eq:quark_level_scalar_hadronic_op} using \pythia, and by incorporating the ChiPT/HLS results described above in a \texttt{FeynRules} model. In the latter approach we include terms with up to three meson fields and up to $\mathcal{O}(1/f_\pi^3)$; the leading decays are into $\pi\pi$, $\pi^0 \eta$, $\pi \pi \pi$, $\rho \pi$, $\omega^0 \pi^0$ and others. 
We note again that the branching fractions are uncertain since multiple operators with unknown Wilson coefficients contribute 
to different channels; this is already apparent for $\chi \to \pi^0 \pi^0$ versus $\chi \to \pi^+ \pi^-$, which are generated by operators in 
Eq.~\eqref{eq:uuuu_hls_1} and~\eqref{eq:uuuu_hls_2}.

\begin{figure}
      \centering
      \includegraphics[width=0.47\textwidth]{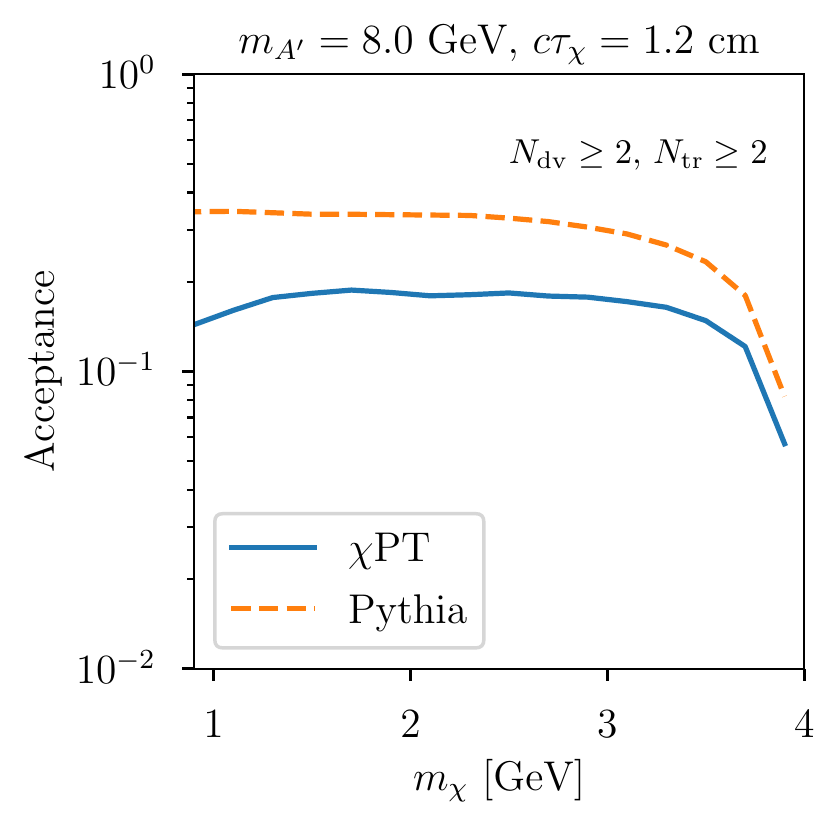}
      \includegraphics[width=0.47\textwidth]{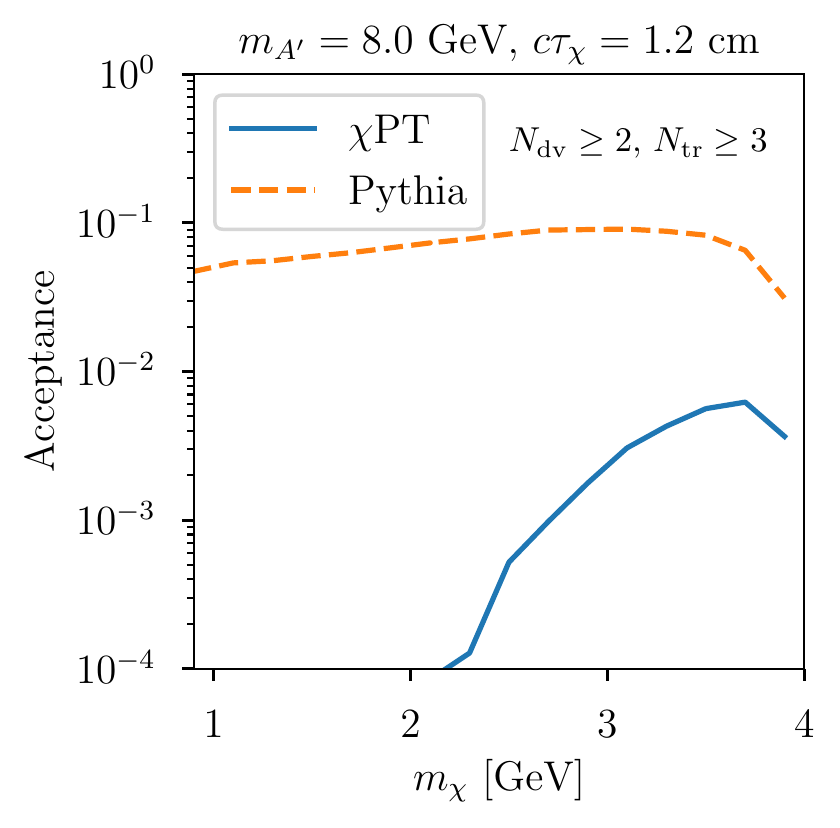}
      \caption{ Comparison of LLP signal acceptances evaluated in two different hadronization models of $\chi \to \bar{u} u\bar{u} u$ for $\mAp = 8\;\GeV$ and $c\tau_\chi \approx 1\;\cm$. 
        In the left (right) panel we show the $\Ndv \geq 2$, $\Ntr \geq 2$ ($\Ndv \geq 2$, $\Ntr \geq 3$) analyses. 
        The solid blue lines show acceptances for final state generated using chiral perturbation theory/hidden local symmetry approach, 
        while the orange dashed lines were obtained using \pythia.\label{fig:acceptance_slices_uuuu_pythia_vs_chipt}}
    \end{figure}

In Fig.~\ref{fig:acceptance_slices_uuuu_pythia_vs_chipt} we show a comparison of the signal acceptance 
for $\chi \to \bar{u}u\bar{u}u$ hadronized in \pythia and using the ChiPT/HLS approach outlined above; 
we focus on a single benchmark point with $\mAp = 8\;\GeV$ and $c\tau \approx 1\;\cm$. 
The results for the $\Ndv \geq 2$, $\Ntr \geq 2$ ($\Ndv \geq 2$, $\Ntr \geq 3$) analysis are given in the left (right) panel. 
In the $\Ntr\geq 2$ case the signal is dominated by $\chi\to \pi^+\pi^-$ decays and we see that the acceptances agree between 
the two hadronization methods within a factor of $\sim 2$. This agreement can be improved by tuning 
the Wilson coefficients of the operators contributing to different branching fractions; we leave this exercise to future work.
As for semi-leptonic decays, the $\Ntr\geq 3$ exhibits significant differences between the \pythia and ChiPT/HLS approaches, 
with \pythia predicting a significantly better acceptance at lower LLP masses. We hypothesize again that this is due to the 
string fragmentation model employed by \pythia not assigning a large enough phase space suppression for producing multiple 
mesons. Thus, high-multiplicity events are much more common in the \pythia event samples compared to ChiPT/HLS, leading to a significantly 
larger acceptance at small LLP masses. At the same time, our ChiPT treatment doesn't include broad vector resonances beyond the lightest multiplet which (as in Sec.~\ref{sec:vector_ud_operator}), which could lead to an under-prediction of higher multiplicity final states in the ChiPT model.

\bibliographystyle{JHEP}
\bibliography{biblio}
\end{document}